\pdfoutput=1
\documentclass{article}
\usepackage[margin=1.0in]{geometry}

\usepackage[T1]{fontenc}
\usepackage{default}

\usepackage{amsmath,amsthm,amssymb,amsfonts}
\usepackage{bm,mathtools,braket,units,array,comment,cases}
\usepackage{fancyhdr,lastpage,lineno}
\usepackage{algorithm,algpseudocode}
\usepackage{makecell}
\usepackage{accents}
\usepackage{filecontents}
\usepackage[dvipsnames]{xcolor}
\usepackage{tikz}
\usepackage[noadjust]{cite}
\usetikzlibrary{bayesnet}
\usepackage[hidelinks]{hyperref}
\usepackage{soul}
\usepackage[normalem]{ulem}

\newcommand\blfootnote[1]{%
	\begingroup
	\renewcommand\thefootnote{}\footnote{#1}%
	\addtocounter{footnote}{-1}%
	\endgroup
}

\makeatletter
\let\start@align@nopar\start@align
\let\start@gather@nopar\start@gather
\let\start@multline@nopar\start@multline
\long\def\start@align{\par\start@align@nopar}
\long\def\start@gather{\par\start@gather@nopar}
\long\def\start@multline{\par\start@multline@nopar}
\makeatother
\algnewcommand{\Inputs}[1]{
	\State \textbf{Inputs:}
	\Statex \hspace*{\algorithmicindent}\parbox[t]{0.95\linewidth}{\raggedright #1}
}
\algnewcommand{\Initialize}[1]{%
	\State \textbf{Initialize:}
	\Statex \hspace*{\algorithmicindent}\parbox[t]{0.95\linewidth}{\raggedright #1}
}

\def\E{\mathbb{E}}
\def\R{\mathbb{R}}
\def\N{\mathcal{N}}

\def\xt{\bm{x}}
\def\xt1{\bm{x}^{(t+1)}}
\def\g{\bm{\gamma}}
\def\G{\bm{\Gamma}}
\def\P{\bm{\Phi}}
\def\e{\bm{e}}

\def\x{\bm{x}}
\def\y{\bm{y}}
\def\z{\bm{z}}

\def\D{\bm{D}}

\def\xsbl{\widehat{\bm{x}}_{\mathrm{SBL}}}

\def\muxy{\bm{\mu}}
\def\Sxy{\bm{\Sigma}}
\def\Sy{\bm{C}}

\def\etal{\textit{\etal.}}
\newcommand{\norm}[1]{\left\|#1\right\|}
\newcommand{\abs}[1]{\left|#1\right|}

\def\noisevar{\lambda}

\title{Sparse Bayesian Learning with Dynamic Filtering for Inference of Time-Varying Sparse Signals}
\author{Matthew R.\ O'Shaughnessy, Mark A.\ Davenport, and Christopher J.\ Rozell}
\date{December 20, 2019}%

\begin{document}

\setstcolor{red}

\maketitle

\blfootnote{M.\ R.\ O'Shaughnessy, M.\ A.\ Davenport, and C.\ J.\ Rozell are with the School of Electrical and Computer Engineering, Georgia Institute of Technology, Atlanta, GA, 30332 USA (e-mail: moshaughnessy6@gatech.edu, mdav@gatech.edu, crozell@gatech.edu). This work was supported in part by NSF grants CCF-1350616 and CCF-1409422, James S.\ McDonnell Foundation grant number 220020399, a gift from the Alfred P.\ Sloan Foundation, and the National Defense Science and Engineering Graduate (NDSEG) Fellowship.}%

\begin{center}
	\textbf{Note:} Single column version of manuscript to appear in \textit{IEEE Transactions on Signal Processing}\footnote{\texttt{https://doi.org/10.1109/TSP.2019.2961229}}.
\end{center}

\begin{abstract}
Many signal processing applications require estimation of time-varying sparse signals, potentially with the knowledge of an imperfect dynamics model. In this paper, we propose an algorithm for dynamic filtering of time-varying sparse signals based on the sparse Bayesian learning (SBL) framework. The key idea underlying the algorithm, termed SBL-DF, is the incorporation of a signal prediction generated from a dynamics model and estimates of previous time steps into the hyperpriors of the SBL probability model. The proposed algorithm is online, robust to imperfect dynamics models (due to the propagation of dynamics information through higher-order statistics), robust to certain undesirable dictionary properties such as coherence (due to properties of the SBL framework), allows the use of arbitrary dynamics models, and requires the tuning of fewer parameters than many other dynamic filtering algorithms do. We also extend the fast marginal likelihood SBL inference procedure to the informative hyperprior setting to create a particularly efficient version of the SBL-DF algorithm. Numerical simulations show that SBL-DF converges much faster and to more accurate solutions than standard SBL and other dynamical filtering algorithms. In particular, we show that SBL-DF outperforms state of the art algorithms when the dictionary contains the challenging coherence and column scaling structure found in many practical applications.
\end{abstract}


\section{Introduction}
\label{sec:introduction}

Many signal processing applications require the reconstruction of \emph{sparse} and \emph{time-varying} signals from undersampled and noisy measurements. For example, in medical imaging it is well-known that MRI and CT images can be accurately represented in a wavelet basis by only a few nonzero coefficients that vary slowly over time due to organ movement \cite{lustig2007sparse,gamper2008compressed}. In direction of arrival problems, targets may be located in only a few spatial locations, and their movement may be governed by a known model \cite{malioutov2005sparse}. The signals in each of these applications are both sparse (i.e., can be represented using only a few elements in some basis or dictionary) and vary over time, perhaps according to some imperfectly known dynamics model.

When recovering \emph{static} sparse signals, incorporating a signal model that exploits this low-dimensional structure can result in dramatic improvements in estimation accuracy, allowing sparse signals to be reconstructed from extremely underdetermined systems \cite{candes2006robust,donoho2006compressed,candes2006compressive}. In dynamic settings, we would like to be able to additionally exploit the a priori knowledge that either the signal changes only slowly over time, or according to a known (but potentially imperfect or noisy) dynamics model. Many effective algorithms for improving inference using dynamics information have been developed; a smaller literature has developed algorithms that exploit both dynamics information and the low-dimensional structure of signals. These algorithms are reviewed in detail in Section \ref{sec:background-sparse-dynamic-filtering}.

Most of these methods for exploiting both sparsity and dynamics information do so by extending variants of $\ell_1$-minimization algorithms. In this paper, however, we consider a sparsity-based dynamic filtering algorithm that makes use of a sparse recovery procedure based on a hierarchical probabilistic model called \emph{sparse Bayesian learning} (SBL) \cite{tipping2001sparse,wipf2004sparse}. We do so for several reasons. First, in many cases the sparse Bayesian learning algorithm has been shown to produce more accurate estimates than $\ell_1$-based sparse recovery methods \cite{wipf2011latent,wipf2010iterative} \cite[Sec.~I-C]{zhang2012sparse}. Second, the SBL procedure has been shown to be robust to undesirable dictionary structure such as the combination of high coherence\footnote{We say that a dictionary $\P$ has high \emph{coherence} if the normalized inner product of its columns, $\underset{i,j}{\max} \frac{\abs{\P_i^T \P_j}}{\lVert\P_i\rVert \lVert \P_j \rVert}$, is large. Intuitively, this makes reconstruction challenging because it is difficult to differentiate measurements of $x_i$ and $x_j$, particularly in the presence of measurement noise \cite{donoho2003optimally}.} and diverse column magnitudes \cite{wipf2011sparse}, structure that violates sufficient conditions for $\ell_1$-based sparse estimation algorithms but that nonetheless commonly appears in real-world applications such as imaging \cite{wipf2010robust}. Third, many applications make use of the SBL framework because its probabilistic framework allows other application-specific priors or constraints to be incorporated into the inference process in a simple and interpretable way (e.g., \cite{nannuru2019sparse}). Finally, the SBL inference procedure allows the noise variance (sparsity) parameter to be learned automatically, reducing the need for manual parameter tuning compared with similar algorithms.

In this paper we develop an SBL-based algorithm for the online recovery of sparse time-varying signals. Our SBL with dynamic filtering (SBL-DF) algorithm replaces the uninformative hyperpriors typically used by the SBL probability model with informative hyperpriors that have parameters set using a dynamics-based signal prediction. This strategy allows our algorithm to incorporate dynamics information in a manner that is particularly robust to noise or modeling error, utilize arbitrary dynamics models, and inherit the beneficial properties of the SBL framework (particularly its robustness to adversely structured dictionaries).

Mathematically, we consider the problem of recovering a signal $\x^{(t)} \in \R^N$ at times $t = 1, \dots, L$ that evolves according to the Markov process
\begin{equation}
	\x^{(t)} = f_t \left( \x^{(t-1)} \right) + \bm{n}^{(t)}, \label{eq:dyn-model}
\end{equation}
where $\x^{(t)}$ is sparse (i.e., $\norm{\x^{(t)}}_0 \ll N$), $f_t ~\colon~ \R^N \rightarrow \R^N$ is the known dynamics model that describes the evolution of the signal from time $t-1$ to time $t$, and $\bm{n}^{(t)}$ represents the error in the dynamics model (referred to as \emph{innovations}). Rather than observe the signal directly, we obtain noisy underdetermined measurements $\y^{(t)} \in \R^M$ through the dictionary $\P \in \R^{M \times N}$ (where typically $M \ll N$) according to the model\footnote{Although we consider $\x$ to be sparse in the canonical domain for clarity of notation, our approach can easily be extended to the case where $\x$ is sparse in some other basis or dictionary. Similarly, although we consider $\P$ to be fixed for all $t$ for simplicity, it can easily be extended to be different for each time $t$.}
\begin{equation}
	\y^{(t)} = \P \x^{(t)} + \e^{(t)} \label{eq:meas-model}
\end{equation}
where $\e^{(t)} \in \R^M$ represents the measurement noise at time $t$.

After reviewing the SBL procedure and existing methods for dynamic filtering of sparse signals in Section \ref{sec:background}, we present our main contributions in Section \ref{sec:sbldf}: the SBL-DF algorithm in Section \ref{sec:sbldf-algorithm}, an investigation of how some properties of static SBL extend to SBL-DF in Section \ref{sec:sbldf-perspectives}, and a fast version of the algorithm using an extension of the fast marginal likelihood procedure \cite{tipping2003fast} in Section \ref{sec:sbldf-fml}. In Section \ref{sec:simulations}, we provide experimental validation of the algorithm and explore how its performance compares to state of the art algorithms in various settings. We conclude in Section \ref{sec:discussion} with a brief discussion of avenues for future work.

\section{Background and Related Work}
\label{sec:background}

\subsection{Sparse Bayesian learning (SBL)}
\label{sec:background-sbl}

\subsubsection{Probability model}
The SBL model makes the usual assumption that measurements are corrupted by i.i.d.\ Gaussian measurement noise, $\e \sim \mathcal{N}(0, \noisevar \bm{I})$, yielding the likelihood
\begin{equation*}
	p(\y \vert \x, \noisevar) = \mathcal{N}(\P \x, \noisevar \bm{I}).
\end{equation*}

A zero-mean Gaussian prior is placed on each element of $\x$,
\begin{equation*}
	p(x_i \vert \gamma_i) = \mathcal{N}_{x_i}\left(0,\gamma_i\right),
\end{equation*}
with variances parameterized by hyperparameters $\g \in \R^N$.

This Gaussian prior does not have the high kurtosis of distributions known to encourage sparsity. Instead, the SBL procedure encourages sparse solutions by choosing appropriate values for $\g$ based on the observed data $\y$. Intuitively, if a small variance $\gamma_i$ is chosen, the probability mass of the prior $p(x_i \vert \gamma_i)$ is concentrated near zero, so significant evidence from the measurements will be required to make $x_i$ nonzero. Conversely, a large variance $\gamma_i$ makes the prior very wide, giving the model more flexibility to select nonzero $x_i$. The specific method for accomplishing this automatic variable selection is described in the next section.

To complete the probability model, a (conjugate) inverse gamma hyperprior is placed on each variance $\gamma_i$ and $\lambda$,
\begin{equation*}
p(\gamma_i) = \mathcal{IG}_{\gamma_i}(a_i,b_i) = \frac{b_i^{a_i}}{\Gamma(a_i)} \gamma_i^{-a_i-1} e^{-b_i \gamma_i^{-1}}
\label{eq:sbl-hyperprior-gamma}
\end{equation*}
and
\begin{equation*}
p(\noisevar) = \mathcal{IG}_{\noisevar}(c,d) = \frac{d^c}{\Gamma(c)} \noisevar^{-c-1} e^{-d\noisevar^{-1}},
\end{equation*}
where $\left\{a_i\right\}$ and $c$ are \emph{shape} parameters, $\left\{b_i\right\}$ and $d$ are \emph{scale} parameters, and $\Gamma(z) = \int_0^{\infty} x^{z-1} e^{-x} dx$. In practice, the parameters of these inverse gamma hyperpriors are typically selected to make the hyperpriors either \emph{uninformative} (i.e., flat) by setting $a_i = b_i = c = d = 0 ~\forall~ i$ or \emph{scale-invariant} (i.e., Jeffreys) by setting $a_i = c = 1,~ b_i = d = 0 ~\forall~ i$ \cite{tipping2001sparse}, \cite[Sec.~2.2.3]{buchgraber2013variational}.



\subsubsection{Inference}
Computing the estimate $\xsbl$ from the SBL probability model requires the calculation of the posterior $p(\x, \g, \noisevar \vert \y)$. Because this quantity cannot be computed analytically, the SBL procedure proceeds by decomposing the full posterior as
\begin{equation}
	p(\x, \g, \noisevar \lvert \y) = p(\x \vert \y, \g, \noisevar) ~ p(\g, \noisevar \vert \y),
	\label{eq:sbl-posterior-decomp}
\end{equation}
where we call $p(\x \vert \y, \g, \noisevar)$ the \emph{source posterior} and $p(\g, \noisevar \vert \y)$ the \emph{hyperparameter posterior}.

The source posterior $p(\x \vert \y, \g, \noisevar)$ admits the closed form expression $p(\x \vert \y, \g, \noisevar) = \mathcal{N}(\muxy,\Sxy)$, where, denoting $\G = \mathrm{diag}(\g)$,
\begin{equation}
	\Sxy = \left( \G^{-1} + \noisevar^{-1} \P^T \P \right)^{-1} \quad\text{and}\quad \muxy = \noisevar^{-1} \Sxy \P^T \y.
	\label{eq:sbl-posterior-covariance-mean}
\end{equation}

This expression allows us to calculate the point estimate $\xsbl = \E_{\x} \left[ p(\x \vert \y,\g,\noisevar) \right] = \muxy$ once the parameters $\g$ and $\lambda$ have been estimated. These parameters are determined from the observations $\y$ by approximating the hyperparameter posterior in (\ref{eq:sbl-posterior-decomp}) by its mode (i.e., $p(\g, \noisevar \vert \y) \approx \delta(\g^*, \noisevar^*)$), which is determined by maximizing the \emph{evidence for the measurements} $p(\y \vert \g, \noisevar)$. Because the evidence is a convolution of Gaussians, it can be computed from the quantities defined in the probability model as $p(\y\vert\g,\noisevar) = \int p(\y \vert \x, \noisevar) p(\x \vert \g) d\x = \N_{\y}(\bm{0},\Sy)$, where
\begin{equation}
	\Sy = \noisevar \bm{I} + \P \G \P^T. \label{eq:sbl-marginal-likelihood-covariance}
\end{equation}

The maximization of the evidence $p(\y\vert\g,\noisevar)$, which is often referred to as \emph{type-II maximum likelihood}, intuitively represents selecting the hyperparameters that are best supported by the observed data $\y$. The SBL procedure maximizes the marginal likelihood and hyperparameters over a log scale\footnote{Maximizing the hyperparameters over a log scale results in Jeffreys scale invariant hyperpriors; see \cite[Appx.~A]{tipping2001sparse}, \cite[Sec.~2.2.2---2.2.3]{buchgraber2013variational}, and \cite{wipf2003perspectives} for more details.}. Placing an uninformative prior on the noise variance $\noisevar$ (i.e., $c = d = 0$) results in the nonconvex negative marginal log-likelihood
\begin{align}
	\hspace{-0.3em}\ell(\g,\noisevar) = -\log \left[ p(\y \vert \log \g, \log \noisevar) \prod_{i=1}^N p(\log \gamma_i) p(\log \noisevar) \right] \notag \\
	\propto \log \left| \Sy \right| + \y^T \Sy^{-1} \y - 2 \sum_{i=1}^N \left( a_i \log \gamma_i^{-1} - b_i \gamma_i^{-1} \right).
	\label{eq:sbl-objective}
\end{align}
Note that when uninformative priors are also placed on the $\gamma_i$ (i.e., $a_i = b_i = 0$ for all $i$), as is the case in the traditional SBL algorithm, the cost function is simply $\ell_{\mathrm{uninf}}(\g,\lambda) = \log \abs{\Sy} + \y^T \Sy^{-1} \y$.

Several methods have been used in the SBL literature to iteratively minimize the cost function (\ref{eq:sbl-objective}). The commonly used expectation-maximization procedure can be derived by treating $\x$ as the hidden variable, resulting in the iterative updates \cite{tipping2001sparse,wipf2004sparse}
\begin{equation}
	\label{eq:sbl-em-gamma}
	\gamma_i^{(k+1)} = \frac{\left(\Sxy\right)_{ii} + \bm{\mu}_i^2 + 2b_i}{1 + 2a_i}
\end{equation}
and
\begin{equation}
	\label{eq:sbl-em-s2}
	\lambda^{(k+1)} = \frac{\norm{\y - \P \muxy}_2^2 + \mathrm{Tr}\left[\P^T\P\Sxy\right]}{M}.
\end{equation}

During the EM procedure, many variances $\gamma_i$ become very small, indicating that the corresponding element $x_i$ is zero. To avoid numerical instability, when any $\gamma_i$ falls below some small threshold $\tau$ (e.g., $\tau = 10^{-12}$), the corresponding $\gamma_i$ and dictionary column $\P_i$ are pruned from the model and $x_i$ is fixed to zero. We denote the set of unpruned indices by $\mathcal{T} = \left\{i \colon \gamma_i \geq \tau \right\}$. An advantageous property of performing this pruning is that each EM iteration (whose cost is dominated by the $O(N^3)$ matrix inverse required to recompute $\Sxy$) becomes significantly faster as more elements are pruned from the model, so in practice larger values of $\tau$ (e.g., $\tau = 10^{-4}$) are sometimes used.

\subsection{Dynamic filtering of sparse signals}
\label{sec:background-sparse-dynamic-filtering}

The Kalman filter (KF) \cite{kalman1960new} is the foundational algorithm for tracking time-varying signals. By assuming the dynamics model is linear, the states have Markovian structure, and the state evolution and observations are corrupted by Gaussian noise, the KF admits optimal closed-form expressions that recursively estimate the state at each time step. Extensions of the KF have also relaxed the linear and Gaussian assumptions by linearizing nonlinear dynamics models \cite{rauch1965maximum} or representing the state distribution using carefully chosen points \cite{julier1997new,wan2000unscented}. Further, although the KF framework is causal, Kalman \emph{smoothing} algorithms (e.g., \cite{rauch1965maximum}) refine the estimate of previous time steps using the complete signal estimate.

However, these methods do not exploit the low-dimensional (sparse) structure present in many applications, which can significantly improve estimation accuracy. One body of work exploits sparse structure with the assumption that the signal support is constant or changing only slowly by incorporating a group $\ell_1$ penalty into the KF \cite{angelosante2009lassokalman}, estimating sparse outliers in the measurements and dynamics model jointly with the states \cite{farahmand2011doubly}, or using support estimation as a preprocessing step for the KF \cite{vaswani2010lscsresidual,vaswani2010modifiedcs}. Other methods exploit the sparsity of time-varying signals even when the support is varying rapidly by modifying the KF measurement model to include ``pseudo-measurements'' that have the effect of encouraging sparsity, inserting a modified orthogonal matching pursuit procedure into the KF procedure \cite{zachariah2012dynamic}, extending the Kalman smoothing framework to use a broader class of penalty functions that promote sparsity \cite{carmi2010methods,aravkin2013sparse}, or modifying the $\ell_1$ homotopy algorithm to use the KF framework \cite{asif2011estimation} or sliding window-based processing \cite{asif2014sparse}.

Another family of algorithms start from an optimization perspective, augmenting the least-squares objective with $\ell_1$ terms that either directly encourage the signal and/or dynamics innovations to be sparse \cite{charles2011sparsity} or encourage the signal to be close to a dynamics-based estimate using $\ell_2$ \cite[Sec.~3]{charles2016dynamic} or optimal transport-inspired \cite{bertrand2018sparse} penalties.

The $\ell_1$ methods most similar to our proposed algorithm are methods that model the time-varying signal using a probabilistic framework by setting the priors of future time steps using dynamics information. In \cite{mecklenbrauker2013sequential}, a weighted $\ell_1$ penalty is used with weights set using the previous signal state and a probabilistic dynamics model. In the RWL1-DF algorithm of \cite{charles2016dynamic}, dynamics information is incorporated in a robust manner by using a dynamics-based signal prediction to set the hyperpriors of the reweighted $\ell_1$ estimator \cite{candes2008enhancing,garrigues2010group}. While these methods use probabilistic models based on the weighted or reweighted $\ell_1$ estimator, our SBL-DF algorithm propagates dynamics information through the hyperpriors of the SBL probability model, giving the resulting algorithm the same advantageous properties that SBL has been shown to have.

\subsection{Dynamic filtering of sparse signals using SBL}
\label{sec:background-sbl-dynamic-filtering}

In contrast to the potpourri of algorithms for dynamic filtering of sparse signals using standard $\ell_1$-based methods, relatively few algorithms exist for incorporating dynamics information into the SBL framework --- despite its widespread use in practical applications.

SBL was initially extended to the offline recovery of dynamic sparse signals with constant support in the MSBL algorithm by assigning a single variance hyperparameter to each row of the matrix $\bm{X} = \left[ \x^{(1)}, \x^{(2)}, \dots, \x^{(L)} \right]$ and jointly recovering the entire sequence of signals \cite{wipf2007empirical}. The TSBL algorithm improves this method by incorporating temporal correlation information into the SBL probability model \cite{zhang2011sparse}; an efficient online version is derived in \cite{joseph2017noniterative}. These methods, however, assume that the signal support is constant over the entire time interval.

The constant support requirement is relaxed in \cite{zhang2011exploiting} by reconstructing the signal over a small sliding window on which the signal support is assumed to be approximately stationary. A similar strategy is used in \cite{buchgraber2011slidingwindow}, which reconstructs the signal on each time window using an efficient variational SBL algorithm \cite{shutin2011fast} rather than TSBL. The sliding window approach is improved in \cite{wijewardhana2017bayesian}, which designs windows using a lapped orthogonal transform (similar to \cite{asif2014sparse}) to avoid blocking artifacts and derives an efficient SBL algorithm to add and remove measurements based on the fast marginal likelihood updates of \cite{tipping2003fast}. These methods remove the constant support assumption of MSBL and TSBL, but do not allow the use of a dynamics model and are ill-suited to applications where the signal is changing rapidly.

Echoing similar developments for the $\ell_1$ case, \cite{karseras2013tracking,filos2013tracking} incorporate the SBL algorithm into the Kalman framework by replacing the Kalman filter's Gaussian prior on the dynamics innovations with the Gaussian-inverse Gamma prior of the SBL framework, then performing SBL inference during each time step. Unlike previous methods, this algorithm allows online recovery of time-varying signals using SBL, but in contrast to our method, assumes the dynamics innovations are sparse rather than the signal itself. This sparse-innovations signal model can result in non-sparse states, and is therefore not the appropriate signal model for many applications.

Most similar to our proposed algorithm are methods that incorporate a dynamics-based signal prediction into the SBL probability model or objective function. In \cite{fang2015support,shekaramiz2016sparse}, an estimate of the signal support is incorporated into the SBL probability model by adding a third-level hyperprior, then performing inference using a variational procedure. Although incorporating the support estimate into hyperpriors in this way makes the model more robust than methods that rigidly enforce (potentially erroneous) support estimates do, the method does not differentiate between small and large nonzero coefficients in the dynamics estimate, and requires an expensive variational procedure to perform inference on the more complex resulting model. In \cite{wang2014exploiting}, the SBL framework is modified by replacing the hyperprior on the variance parameters with an improper Gaussian parameterized by the previous value of the signal. Inference in this modified model is performed on the signal innovations at each time step, reducing the strength of SBL's sparsity encouraging mechanism for elements that are large in previous time steps. Unlike the sparse-states signal model we consider, this sparse-innovations model can lead to non-sparse states. Further, none of these methods allow the use of a dynamics model to improve inference.

\section{Dynamic Filtering via Sparse Bayesian Learning (SBL-DF)}
\label{sec:sbldf}

\subsection{The SBL-DF algorithm}
\label{sec:sbldf-algorithm}

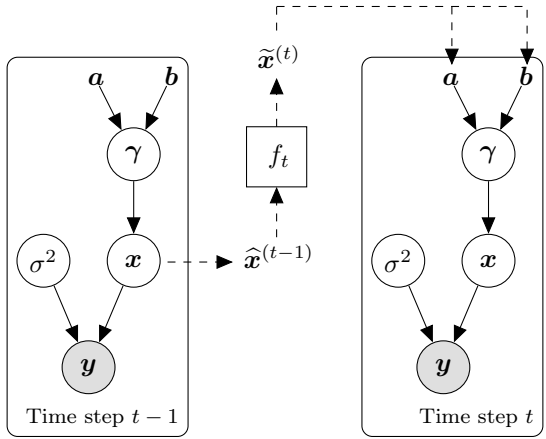
\begin{figure}[h!]
	\centering
	\begin{tikzpicture}
		\node[obs]
			(y1) {$\y$};
		\node[latent, above=of y1, xshift=-0.6cm, yshift=-0.3cm]
			(s21) {$\sigma^2$};
		\node[latent, above=of y1, xshift=0.6cm, yshift=-0.3cm]
			(x1) {$\x$};
		\node[latent, above=of x1, yshift=-0.3cm]
			(g1) {$\g$};
		\node[const, above=of g1, xshift=-0.5cm, yshift=-0.45cm]
			(a1) {$\bm{a}$};
		\node[const, above=of g1, xshift=0.5cm, yshift=-0.45cm]
			(b1) {$\bm{b}$};
		
		\node[obs, right=of y1, xshift=3cm]
			(y2) {$\y$};
		\node[latent, above=of y2, xshift=-0.6cm, yshift=-0.3cm]
			(s22) {$\sigma^2$};
		\node[latent, above=of y2, xshift=0.6cm, yshift=-0.3cm]
			(x2) {$\x$};
		\node[latent, above=of x2, yshift=-0.3cm]
			(g2) {$\g$};
		\node[const, above=of g2, xshift=-0.5cm, yshift=-0.45cm]
			(a2) {$\bm{a}$};
		\node[const, above=of g2, xshift=0.5cm, yshift=-0.45cm]
			(b2) {$\bm{b}$};
		
		\node[const, right=of x1, xshift=0.1cm, yshift=0.065cm]
			(xhat1) {$\widehat{\bm{x}}^{(t-1)}$};
		\node[const, above=of xhat1, xshift=0cm, yshift=0cm]
			(f) {$f_t$};
		\node[const, above=of f, xshift=0cm, yshift=0cm]
			(xtilde) {$\widetilde{\x}^{(t)}$};
		\draw(2.1,2.4) rectangle (2.9,3.2);
		
		\edge {x1,s21} {y1};
		\edge {g1} {x1};
		\edge {a1,b1} {g1};
		\edge {x2,s22} {y2};
		\edge {g2} {x2};
		\edge {a2,b2} {g2};
		\draw[arrows={-triangle 45},dashed](1.05,1.4)--(1.92,1.4); 
		\draw[arrows={-triangle 45},dashed](2.5,1.73)--(2.5,2.4); 
		\draw[arrows={-triangle 45},dashed](2.5,3.2)--(2.5,3.85); 
		\draw[dashed](2.5,4.48)--(2.5,4.8); 
		\draw[arrows={-triangle 45},dashed](2.5,4.8)--(5.83,4.8)--(5.83,4.02);
		\draw[arrows={-triangle 45},dashed](4.83,4.8)--(4.83,4.02);
		\plate {t1} {(y1)(x1)(s21)(g1)(a1)(b1)} {Time step $t-1$};
		\plate {t2} {(y2)(x2)(s22)(g2)(a2)(b2)} {Time step $t$};
	\end{tikzpicture}
	\caption{Graphical model representation of SBL-DF. The dotted arrow represents that the estimate $\widehat{\x}^{(t-1)}$ is propagated through the dynamics model $f_t$ and into the second-order statistics of the next time step, $\bm{a}^{(t)}$ and $\bm{b}^{(t)}$.}
	\label{fig:sbldf-graphical-model}
\end{figure}

In this section, we introduce our algorithm for estimating time-varying sparse signals using SBL. The key insight of our approach is that the estimate of $\x^{(t)}$ can be improved in a robust manner by injecting information from the estimate of the previous time step and a dynamics model into the hyperparameters of the SBL probability model. A graphical depiction of this strategy is shown in Figure \ref{fig:sbldf-graphical-model}.

The dynamics-based prediction of the state at time $t$, $\widetilde{\x}^{(t)}$, is obtained by propagating $\widehat{\x}^{(t-1)}$ through the dynamics model $f_t \colon \R^N \rightarrow \R^N$. In applications where $f_t$ is known, our prediction is simply $\widetilde{\x}^{(t)} = f_t(\x^{(t-1)})$. Note that, unlike many algorithms based on the Kalman filter, we do not require the dynamics function to be linear. In other applications the dynamics model is unknown, but we expect that $\x^{(t)} \approx \x^{(t-1)}$ (i.e., $\x$ varies only slowly over time). In this case, we simply use either the identity dynamics function $f_t(\x^{(t)}) = \x^{(t-1)}$ or select an $f_t$ that reflects a ``distribution'' on $\x^{(t)}$ based on $\x^{(t-1)}$ (for instance, a Gaussian blurring kernel centered on $\x^{(t-1)}$). In either case, because an important advantage of our method for propagating dynamics information through hyperpriors is the robustness of the resulting model, the use of an inexact dynamics model still provides good results. For example, we show in Section \ref{sec:simulations} that the identity dynamics function works well even when $\x$ is slowly time-varying. Note that although our model does not explicitly estimate or exploit temporal correlation structure as batch estimation procedures such as \cite{zhang2011sparse} do, if a priori correlation information is known it can be incorporated into the dynamics model $f_t$.

We next describe our specific method for mapping the dynamics-based prediction $\widetilde{\x}^{(t)} = f_t \left( \widehat{\x}^{(t-1)} \right)$ to the hyperparameters of the SBL probability model. For clarity of notation, we drop the superscript $t$, denoting $\widehat{\x}^{(t)}$ by $\widehat{\x}$ and $\widetilde{\x}^{(t)}$ by $\widetilde{\x}$.

First, we note that the SBL objective (\ref{eq:sbl-objective}) can be decomposed as $\ell\left(\g,\noisevar\right) = \ell_{\mathrm{uninf}}\left(\g,\noisevar\right) + \ell_{\mathrm{dyn}}\left(\g,\noisevar\right)$, where
\begin{equation}
	\ell_{\mathrm{dyn}}\left(\g,\noisevar\right) = 2\sum_{i}\left(-a_i\log\gamma_i^{-1}+b_i\gamma_i^{-1}\right) \label{eq:sbl-objective-dyn}
\end{equation}
represents the dynamics-based portion of the objective that we will use to encourage fidelity to the dynamics estimate. The component $\ell_{\mathrm{uninf}}\left(\g,\noisevar\right)$, which promote sparsity and fidelity to the measurements, is exactly the standard uninformative SBL objective. Intuitively, the overall magnitudes of $a_i$ and $b_i$ can be interpreted as controlling how strongly $\ell_{\mathrm{dyn}}\left(\g,\noisevar\right)$ is weighed in the objective, and the relative magnitudes of $a_i$ and $b_i$ can be interpreted as controlling the specific value of $\gamma_i$ that $\ell_{\mathrm{dyn}}\left(\g,\noisevar\right)$ encourages.

We select the hyperparameter values $\left\{a_i,b_i\right\}$ so that the $\g$ minimizing $\ell_{\mathrm{dyn}}\left(\g,\noisevar\right)$ coincides with the $\g$ minimizing the expected difference between the SBL-DF signal estimate $\widehat{\x}$ and the dynamics-based prediction $\widetilde{\x}$. This selection of $\left\{a_i,b_i\right\}$ is accomplished in two steps. First, we determine the variance parameters as
\begin{equation}
	\g_{\mathrm{dyn}} = \underset{\g}{\arg\min}~\E\left[\norm{\widehat{\x}-\widetilde{\x}}_2^2\right].
	\label{eq:sbldf-gamma-dyn}
\end{equation}
A closed-form expression for (\ref{eq:sbldf-gamma-dyn}) cannot be found in general, so we make the simplifying assumption that $\P^T\P$ is diagonal (i.e., $\P$ has orthogonal columns, but is not necessarily orthonormal). We have found that this approximation works well in practice even when the columns of $\P$ are not orthogonal. Defining $g_i = \P_i^T\P_i$ and taking the derivative (with respect to $\g^{-1}$ for simplicity), we have
\begin{align*}
	\frac{\partial}{\partial\g^{-1}} &\E \left[ \norm{\widehat{\x}-\widetilde{\x}}_2^2 \right] \\ =&\frac{\partial}{\partial\g^{-1}} \E \left[ \norm{\left(\lambda\G^{-1}+\P^T\P\right)^{-1}\P^T\y-\widetilde{\x}}_2^2 \right] \\
	=& \sum_{i=1}^N \frac{\partial}{\partial\gamma_i^{-1}} \left[ \left(\frac{g_i}{\lambda\g_i^{-1}+g_i} - 1\right)^2 \widetilde{x}_i^2 + \noisevar \frac{g_i}{\left(\lambda\gamma_i^{-1}+g_i\right)^2} \right] \\
	=& \sum_{i=1}^N 2\lambda \frac{\lambda\gamma_i^{-1}b_i\widetilde{x}_i^2 - \noisevar b_i}{\left(\lambda\gamma_i^{-1}+b_i\right)^3},
\end{align*}
where the first equality follows from substituting the SBL posterior mean in (\ref{eq:sbl-posterior-covariance-mean}) for $\widehat{\x}$ and then the measurement model (\ref{eq:meas-model}) for $\y$. Setting to zero and solving for $\gamma_i$ yields
\begin{equation}
	\gamma_{\mathrm{dyn},i} = \widetilde{x}_i^2 \label{eq:sbldf-map-g-x}
\end{equation}
for each component. This $\g$ represents the variance parameters that would be optimal if the dynamics estimate $\widetilde{\x}$ was perfect, and is therefore the value of $\g$ that we would like the dynamics-based terms of the informative SBL objective $\ell(\g,\noisevar)$ to encourage.

We now note that for any fixed $a_i, b_i$, the $\gamma_i$ minimizing $\ell_{\mathrm{dyn}}$ is
\begin{equation}
	\gamma_i^* = \frac{b_i}{a_i}. \label{eq:sbldf-map-g-ab}
\end{equation}

Equating the maps (\ref{eq:sbldf-map-g-x}) and (\ref{eq:sbldf-map-g-ab}) gives the rule $b_i / a_i = \widetilde{x}_i^2$. Incorporating a multiplicative ``trade-off'' parameter gives the dynamics mapping formula
\begin{equation}
	a_i = \xi \quad\text{and}\quad b_i = \xi~ \widetilde{x}_i^{~2}, \label{eq:sbldf-dynamics-mapping}
\end{equation}
where the parameter $\xi$ represents how much weight the dynamics-based prediction is assigned in the evidence maximization procedure. Larger values of $\xi$ weigh the dynamics estimate more strongly when selecting $\g$ and therefore reflect greater confidence in the accuracy of the dynamics model; see Section \ref{sec:simulations-dynquality} for details.

Note that although the SBL-DF algorithm as presented runs causally using only a single previous estimate $\widehat{\x}^{(t-1)}$, it can easily be extended to use multiple previous estimates or into a smoothing estimator by setting $a_i^{(t)}$ and $b_i^{(t)}$ based on the estimate of $\x$ at other prior or future time steps.

Algorithm \ref{alg:sbl-df-em} summarizes the complete SBL-DF algorithm using expectation-maximization iterations to perform inference in the informative hyperprior SBL model.

\begin{algorithm}
	\caption{SBL-DF Using Expectation-Maximization}
	\label{alg:sbl-df-em}
	\begin{algorithmic}[1]
		\State{Initialize $\gamma_i = 1$, $a_i = b_i = 0,~i = 1,\dots N$} 
		\For{$t = 1, \dots L$}
		\While{not converged}
		\State Update $\Sxy$ and $\muxy$ using (\ref{eq:sbl-posterior-covariance-mean})
		\State Update $\g$ and $\noisevar$ using (\ref{eq:sbl-em-gamma}--\ref{eq:sbl-em-s2})
		\State Prune $\muxy$ and $\Sxy$ where $\gamma_i < \tau$
		\EndWhile
		\State Calculate and output $\widehat{\x}^{(t)} = \muxy$
		\State Compute $\bm{b}$ for next time step from $\widehat{\x}^{(t)}$ using (\ref{eq:sbldf-dynamics-mapping})
		\EndFor 
	\end{algorithmic}
\end{algorithm}

\subsection{Implications of informative hyperprior SBL}
\label{sec:sbldf-perspectives}

In contrast to the standard uninformative hyperprior SBL model that has enjoyed success in the literature, the SBL-DF algorithm places informative hyperpriors on the variances $\g$. In this section, we investigate some key properties of the objective function that results from this model and their effects on the SBL-DF algorithm.

\subsubsection{``Effective'' SBL prior}

\begin{figure*}[t]
	\centering
	\begin{tabular}{cc}
		\includegraphics[width=3in]{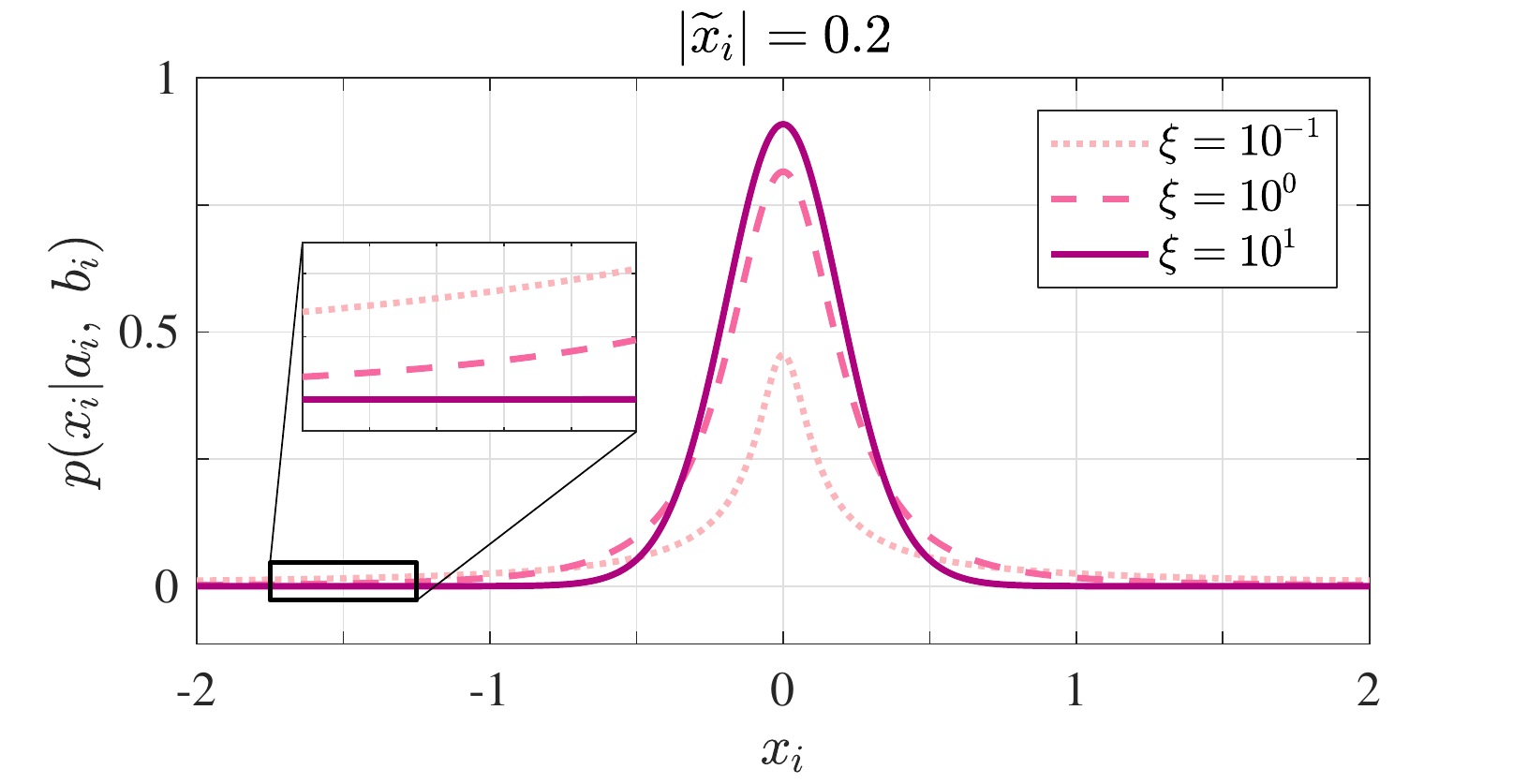} &
		\includegraphics[width=3in]{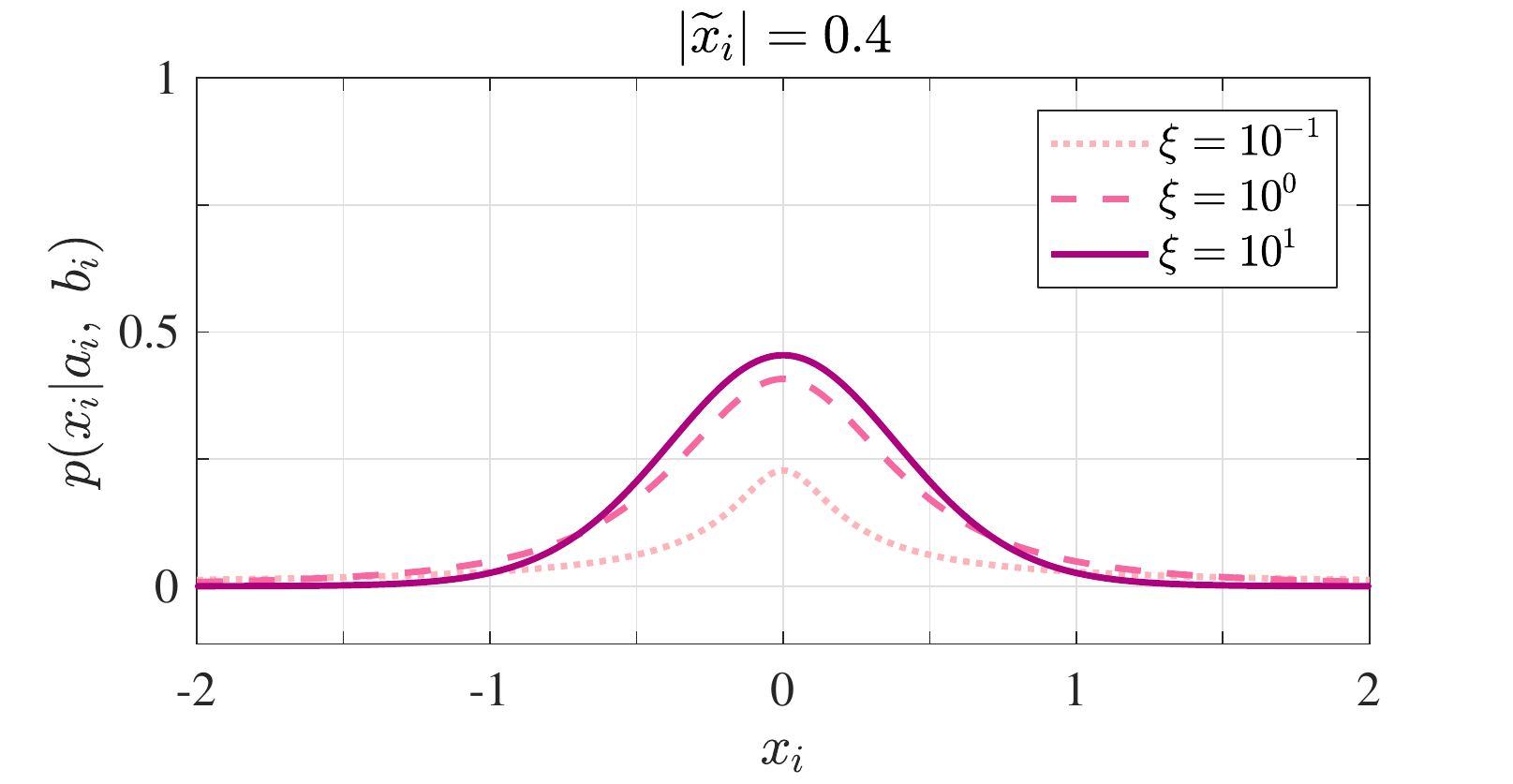} 
	\end{tabular}
	\caption{Student's-$t$ distributions resulting from marginalizing over $\gamma_i$ for SBL hyperparameters set using the rule (\ref{eq:sbldf-dynamics-mapping}). \emph{Left:} The dynamics parameter $\xi$ strengthens or moderates the strength of the prior by changing the height of the peak at $x_i = 0$ and the weight of the tails. As $\xi$ increases, the effective prior becomes more strongly peaked at $x_i = 0$ and its tail probabilities decrease (see left inset). \emph{Right:} When the dynamics estimate magnitude is larger, the prior becomes less tightly peaked, giving the estimator more flexibility to make $x_i$ nonzero.}
	\label{fig:studentt}
\end{figure*}

One method for interpreting the sparsity promoting properties of SBL is to consider the ``effective'' SBL prior obtained by marginalizing out the first layer Gaussian prior $p(\x\vert\g)$. In the SBL probability model, marginalizing out the hyperparameters $\g$ shows us that each $p(x_i)$ is the independent non-standardized Student's-$t$ distribution \cite{tipping2001sparse}
\begin{align*}
	p(x_i \vert a_i, b_i) &= \int_0^{\infty} p(x_i \vert \gamma_i) p(\gamma_i \vert a_i, b_i) d\gamma_i \\
	&= \frac{b_i^{a_i} \Gamma(a_i + \tfrac{1}{2})}{\sqrt{2\pi} \Gamma(a_i)} \left( b_i + \frac{x_i^2}{2} \right)^{-\left(a_i + \tfrac{1}{2}\right)}
\end{align*}
parameterized by location $\mu_i = 0$, scale $\sigma_i = \sqrt{b_i/a_i} = \abs{\widetilde{x}_i}$, and degrees of freedom $\nu_i = 2 a_i = 2\xi$. As demonstrated in \cite{tipping2001sparse}, the high kurtosis of this effective prior intuitively explains why the SBL procedure produces sparse solutions in practice.

From this perspective, we see that SBL-DF's method for propagating dynamics information into the hyperpriors adheres to our intuition that elements with small predicted magnitude should be given an effective prior $p(x_i \vert a_i, b_i)$ tightly peaked about zero, while elements with larger predicted magnitudes should have wider effective priors, which reduces the evidence needed to infer a nonzero value for $x_i^{(t+1)}$. Figure \ref{fig:studentt} shows the effective prior for different values of $\widetilde{x}_i$ and $\xi$. We see that, as desired, the probability mass of the effective prior $p(x_i \vert a_i, b_i)$ increasingly concentrates near zero as $\widetilde{x}_i$ decreases. Meanwhile, as $\xi$ increases the improper hyperprior becomes stronger, increasing the effect of the dynamics estimate on the inferred signal.

\subsubsection{Characterization of the informative hyperprior objective}

\begin{figure}[h!]
	\centering
	\includegraphics[width=3.4in]{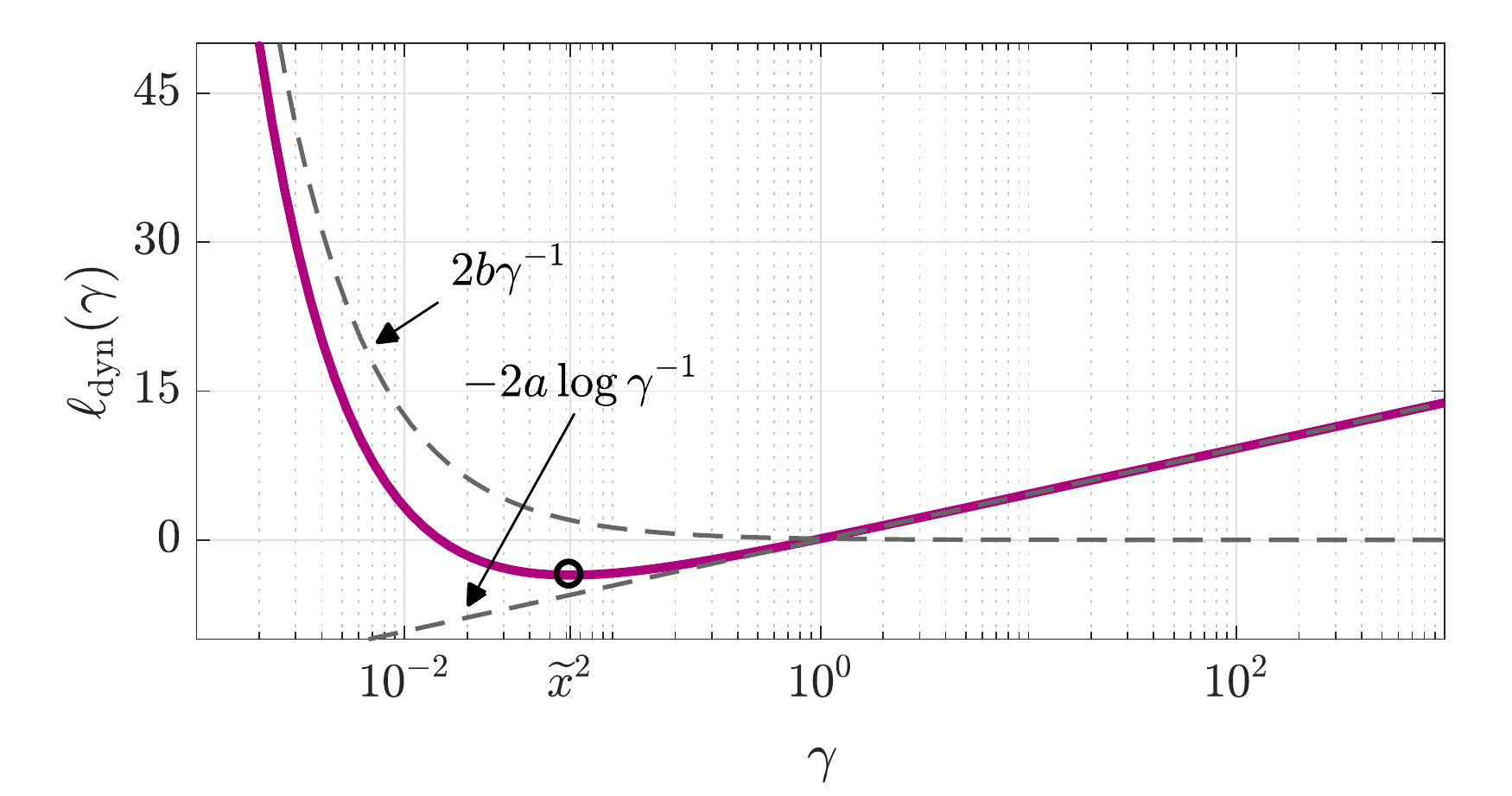}
	\caption{Visualization of portion of the objective resulting from the informative hyperpriors, $\ell_{\mathrm{dyn}}\left(\gamma\right) = -2a\log\gamma^{-1}+2b\gamma^{-1}$, in one dimension for $\widetilde{x} \neq 0$. The term $2b\gamma^{-1} = 2\xi\gamma^{-1}$ penalizes large values of $\gamma$ in a manner not dependent on the signal estimate $\widetilde{x}$. The term $-2a\log\gamma^{-1} = -2\xi\widetilde{x}^2\log\gamma^{-1}$ encourages $\gamma$ to be nonzero when the signal estimate $\widetilde{x}$ has large magnitude.}
	\label{fig:ldyn}
\end{figure}

Figure \ref{fig:ldyn} shows the form of $\ell_{\mathrm{dyn}}\left(\gamma\right)$ in one dimension, demonstrating the contrasting effects of the terms $-2a\log\gamma^{-1}$ and $2b\gamma^{-1}$ on the SBL inference procedure. We observe that while the $-2a\log\gamma^{-1}$ term weakly encourages $\gamma$ to be zero-valued (regardless of the value of $\widetilde{x}$), the $2b\gamma^{-1}$ term encourages $\gamma$ to be nonzero when the magnitude of $\widetilde{x}$ is large. The action of $\ell_{\mathrm{dyn}}\left(\gamma\right)$ in SBL-DF can therefore be interpreted as mitigating SBL's natural sparsity-promoting properties in an amount commensurate with the magnitude of the dynamics-based signal prediction, while more strongly encouraging sparsity when $\widetilde{x}$ is zero-valued.

The minima of the uninformative hyperprior SBL objective $\ell_{\mathrm{uninf}}\left(\g\right)$ are characterized in \cite[Thm.~2]{wipf2004sparse}, which shows that all local minima $\widehat{\x}$ are sparse in the sense that $\norm{\widehat{\x}}_0 < M$. Figure \ref{fig:ldyn} demonstrates that this property may not always be preserved in the informative hyperprior setting when there are many elements where $x_i = 0$ but $\widetilde{x}_i \neq 0$, since when $\widetilde{x}_i$ is nonzero, $2b_i\gamma_i^{-1} = 2\xi\widetilde{x}_i\gamma_i^{-1} \rightarrow \infty$ as $\gamma_i \rightarrow 0$.

\subsubsection{Implementation using reweighted $\ell_1$ iterations}
In \cite{wipf2008new,wipf2010iterative}, Wipf et al.\ examine the properties of the SBL objective by demonstrating that inference can be performed with a reweighted $\ell_1$ minimization procedure (SBL-RWL1) using a modified weight update step. This formulation of the SBL inference procedure elucidates the connection between SBL and $\ell_1$ minimization algorithms, and reveals dictionary dependence in the weight updates of SBL-RWL1 that may be responsible for its superior performance in the presence of adversely structured dictionaries \cite{wipf2011sparse,he2017bayesian}. In this section, we extend the results of \cite{wipf2010iterative} from the uninformative hyperprior setting to the informative hyperprior setting used by SBL-DF, and show how conclusions drawn in literature from uninformative hyperprior SBL-RWL1 \cite{wipf2008new,wipf2010iterative,wipf2011sparse,he2017bayesian,xin2018building} carry over to informative hyperprior SBL-RWL1 and therefore to SBL-DF.

\emph{Proposition 1}: The informative hyperprior SBL objective (\ref{eq:sbl-objective}) can be minimized using the modified reweighted $\ell_1$ iterations
\begin{align}
	z_i^{(k+1)} &= \bm{\phi}_i^T \left(\Sy^{(k)}\right)^{-1} \bm{\phi}_i + \frac{2 a_i}{\gamma_i^{(k)}} \label{eq:rwl1-sbl-overview-z} \\
	\gamma_i^{(k+1)} &= \left(z_i^{(k+1)}\right)^{-1/2} \sqrt{\left(x_i^{(k)}\right)^2 + 2 b_i} \label{eq:rwl1-sbl-overview-g} \\
	\x^{(k+1)} &= \underset{\left\{x_i \colon i \in \mathcal{T} \right\}}{\arg\min}~ \left[\vphantom{\sqrt{x_i^2 + 2 b_i}}\right.\norm{\y-\P\x}_2^2 \notag\\ &\quad\quad+ 2 \lambda \sum_{i \in \mathcal{T}} \left(z_i^{(k+1)}\right)^{1/2} \left.\sqrt{x_i^2 + 2 b_i}~~\right], \label{eq:rwl1-sbl-overview-x}
\end{align}
where $\bm{\phi}_i$ is the $i^{\mathrm{th}}$ column of $\P$, the update to $\z$ is called the \emph{majorization} step and the update to $\x$ is called the \emph{minimization} step. Note that as in the EM procedure a pruning rule is used, fixing $x_i$ to zero and removing column/entry $i$ from $\P$, $\z$, and $\g$ when $\gamma_i \leq \tau$.

The derivation of this procedure closely follows the method of \cite{wipf2008new,wipf2010iterative}, which takes a majorization-minimization approach to minimizing (\ref{eq:sbl-objective}) using a majorizing function cleverly designed to have a form similar to the weighted LASSO objective. Here, we briefly describe the main differences from uninformative hyperprior SBL and their implications; the complete derivation is outlined in Appendix \ref{sec:appendix-rwl1-sbl}.

As in \cite{wipf2010iterative}, the majorizing function is formed as the sum of two upper bounding functions. The concave part of (\ref{eq:sbl-objective}), $\log\abs{\Sy} - 2\sum_ia_i\log\gamma_i^{-1}$, is bounded using its concave conjugate. The addition of the term $-2\sum_ia_i\log\gamma_i^{-1}$ does not change the linear form of this bound, so the effect of the $\{a_i\}$ only appears in the majorization step. The convex part of (\ref{eq:sbl-objective}), $\y^T \Sy^{-1} \y + 2 \sum_i b_i \gamma_i^{-1}$, is bounded by a penalized least-squares term computed by expanding $\y^T \Sy^{-1} \y$ using the Woodbury identity and recognizing the result as the solution to a ridge regression problem; in our case, we add the additional term $2\sum_ib_i\gamma_i^{-1}$ directly into the bound, so the effect of the $\{b_i\}$ appears in the minimization step. The resulting majorization-minimization procedure takes the form of the iterations in Proposition 1, and the complete SBL-DF algorithm using this procedure is listed in Algorithm \ref{alg:sbl-df-rwl1}.

\begin{algorithm}
	\caption{SBL-DF Using Reweighted $\ell_1$ Minimization}
	\label{alg:sbl-df-rwl1}
	\begin{algorithmic}[1]
		\State{Initialize $\gamma_i = 1$, $a_i = b_i = 0,~i = 1,\dots N$} 
		\For{$t = 1, \dots L$}
		\While{not converged}
		\State Update $\z$ and $\g$ using (\ref{eq:rwl1-sbl-overview-z}--\ref{eq:rwl1-sbl-overview-g}) \Comment{Majorize}
		\State Prune where $\gamma_i < \tau$
		\State Update $\x$ using (\ref{eq:rwl1-sbl-overview-x}) \Comment{Minimize}
		\EndWhile
		\State Output $\widehat{\x}^{(t)}$
		\State Compute $\bm{b}$ for next time step from $\widehat{\x}^{(t)}$ using (\ref{eq:sbldf-dynamics-mapping})
		\EndFor 
	\end{algorithmic}
\end{algorithm}

When $b_i = 0$ for all $i$, by considering the $z_i^{1/2}$ as ``weights'' in an $\ell_1$ minimization problem, (\ref{eq:rwl1-sbl-overview-x}) is exactly equivalent to the maximization step in the reweighted $\ell_1$ algorithm \cite{candes2008enhancing}. Further, when $a_i = b_i = 0$ for all $i$, the procedure that iterates between (\ref{eq:rwl1-sbl-overview-z}) and (\ref{eq:rwl1-sbl-overview-x}) is equivalent to the uninformative hyperprior SBL-RWL1 procedure of \cite{wipf2010iterative}.

This procedure facilitates a direct comparison between the SBL procedure and traditional sparsity-encouraging algorithms (see, e.g., \cite{wipf2008new,wipf2010iterative,wipf2011sparse}). In particular, this allows us to verify that some beneficial properties of standard (uninformative hyperprior) SBL still apply when using informative hyperpriors as in SBL-DF. For instance, consider the weight update (\ref{eq:rwl1-sbl-overview-z}). It can be shown \cite{he2017bayesian} that (\ref{eq:rwl1-sbl-overview-z}) can be written as
\begin{align}
	z_i^{(k+1)} &= \underset{\bm{s}}{\min}~ \left[ \frac{1}{\lambda} \norm{\bm{\phi}_i - \P\bm{s}}_2^2 + \sum_{j \in \mathcal{T}}\frac{s_j}{\gamma_j^{(k)}} \right] + \frac{2a_i}{\gamma_i^{(k)}}, \label{eq:rwl1-sbl-z-alt}
\end{align}
where the optimization problem is understood to be performed on the pruned model. The form of (\ref{eq:rwl1-sbl-z-alt}) allows us to verify that two critical properties of standard (uninformative hyperprior) SBL also hold for informative hyperprior SBL and therefore SBL-DF. First, unlike standard RWL1, the weight update is \emph{non-separable}, that is, the value of the $i^{\text{th}}$ weight is dependent on \emph{all} elements in $\x$ through the latent variables $\g$. This contrasts with standard reweighted $\ell_1$, in which the weight on $x_i$ is dependent only on the the single element $x_i$ from the previous iteration. Second, the $\ell_2$ norm term in (\ref{eq:rwl1-sbl-z-alt}) reveals that the weights in SBL-RWL1 depend on the coherence structure of $\P$, despite the inclusion of the nonzero hyperparameter $a_i$. This dependence allows SBL to moderate the penalty applied to coefficients that correspond to highly correlated columns of $\P$ \cite{he2017bayesian}, potentially explaining its superior performance with structured dictionaries.

\subsection{Improving the speed of SBL-DF using fast marginal likelihood}
\label{sec:sbldf-fml}

The EM iterations (\ref{eq:sbl-em-gamma}--\ref{eq:sbl-em-s2}) used in the SBL algorithm to minimize $\ell(\g,\noisevar)$ require an $O(\widetilde{N}^3)$ matrix inversion at each iteration to recompute $\Sxy$, where $\widetilde{N} = \abs{\mathcal{T} = \left\{\gamma_i>\tau\right\}}$ denotes the number of unpruned elements remaining in the model. In \cite{tipping2003fast}, Tipping and Faul propose a more efficient method, the \emph{fast marginal likelihood} (FML) algorithm, for minimizing the uninformative SBL objective $\ell_{\mathrm{uninf}}(\g,\noisevar)$. The key idea of the method in \cite{tipping2003fast} is that the effect of updating an individual coefficient on the state $\Sxy^{-1}$, which is required to calculate the state estimate $\bm{\mu}$, can be computed inexpensively by a rank-one update. Therefore, by updating only a single coefficient at each iteration, we can avoid the full recomputation of $\Sxy^{-1}$ that each EM iteration requires. Moreover, by strategically choosing which coefficient to update at each iteration, convergence can occur with relatively few of these rank-one updates.

However, the FML updates described in \cite{tipping2003fast} require the hyperparameters of the SBL probability model to be uninformative, so the FML procedure cannot directly be applied to the SBL-DF algorithm. In this section, we derive a generalized FML procedure for the informative hyperprior setting (i.e., to minimize $\ell\left(\g,\noisevar\right)$ when $\ell_{\mathrm{dyn}}\left(\g,\noisevar\right) \neq 0$) and adapt it to the SBL-DF algorithm. We show that the FML procedure can be extended to the informative hyperprior SBL model at the cost of a more computationally expensive procedure to choose the best coefficient to update at each iteration, an additional cost that grows as $O(N)$.

The critical fact that enables the FML method is that the covariance matrix $\Sy$ of the marginal likelihood --- and therefore the marginal likelihood $\ell\left(\g,\noisevar\right)$ itself --- is separable in $\g$:
\begin{equation*}
	\Sy = \noisevar \bm{I} + \P \G \P^T = \noisevar \bm{I} + \sum_{i=1}^N \gamma_i \bm{\phi}_i \bm{\phi}_i^T. 
\end{equation*}

This decomposition allows basis vectors to be efficiently added and removed from the model using cheap rank-one updates to $\Sy$. To see this, note that the covariance matrix excluding the $j^{\text{th}}$ element (denoted $\Sy_{-j}$) can be calculated as $\Sy_{-j} = \Sy - \gamma_j\bm{\phi}_j\bm{\phi}_j^T$. Substituting these expressions into the objective (\ref{eq:sbl-objective}) and then separating the terms involving $\gamma_j$ from those involving $\g_{-j}$ and $\noisevar$ shows that the negative log-likelihood (\ref{eq:sbl-objective}) can be decomposed as $\ell(\g,\noisevar) = \ell(\g_{-j}, \noisevar) + \ell(\gamma_j)$ where
\begin{align*}
	\ell(\g_{-j},\lambda) &= \log \left| \bm{C}_{-j} \right| + \y^T \bm{C}_{-j}^{-1} \y \\ &\quad~~ - 2 \sum_{i=1, i \neq j}^N \left( a_i \log \gamma_i^{-1} - b_i \gamma_i^{-1} \right)
\end{align*}
and
\begin{align}
	\ell(\gamma_j) &= \log \left( \gamma_j^{-1} + s_j \right) - \frac{q_j^2}{\gamma_j^{-1} + s_j} \notag \\ &\quad~~ - \left( 2 a_j + 1 \right) \log \gamma_j^{-1} + 2 b_j \gamma_j^{-1}, \label{eq:sbl-fml-lgammaj}
\end{align}
where, following the notation of \cite{tipping2003fast}, $s_j = \bm{\phi}_j^T \bm{C}_{-j}^{-1} \bm{\phi}_j$ and $q_j = \bm{\phi}_j^T \bm{C}_{-j}^{-1} \y$. Critically, this decomposition reveals that finding the value of $\gamma_j$ that minimizes $\ell(\g,\lambda)$ is equivalent to finding the value of $\gamma_j$ that minimizes $\ell(\gamma_j)$. We compute this value, which we denote $\widetilde{\gamma}_j$, by finding the roots of the derivative
\begin{align}
	\frac{\partial \ell(\gamma_j)}{\partial \gamma_j} &= - \frac{1}{\gamma_j^2 s_j + \gamma_j} - \frac{q_j^2}{(1+\gamma_j s_j)^2} \notag \\ &\quad~~ + \left( 2 a_j + 1 \right) \gamma_j^{-1} - 2 b_j \gamma_j^{-2} \notag \\
	&= \frac{2 s_j^{-2}}{\gamma_j^2 \left(\gamma_j+s_j^{-1}\right)^2} \left( c_3 \gamma_j^3 + c_2 \gamma_j^2 + c_1 \gamma_j + c_0 \right), \label{eq:fml-dlgjdgj}
\end{align}
where $c_3 = \left(\tfrac{1}{2}+a_j\right) s_j^2$, $c_2 = \left[\left(\tfrac{1}{2}+2a_j\right)s_j-\tfrac{1}{2}q_j^2-b_js_j^2\right]$, $c_1 = \left(a_j-2b_js_j\right)$, and $c_0 = -b_j$. Roots occur when either $\gamma_j \rightarrow +\infty$ (from both the $\gamma_j$ and $(\gamma_j + s_j^{-1})^2$ terms in the denominator) or when $\gamma_j$ is a root of the cubic expression in parentheses; its roots can be computed analytically or by calculating the eigenvalues of its companion matrix. See Appendix \ref{sec:appendix-fml} for details.

We immediately remove nonreal fixed points $\widetilde{\gamma}_j$, and of the remaining points, keep only the $\widetilde{\gamma}_j$ corresponding to local \emph{minima} of $\ell(\gamma_j)$ by removing roots where the second derivative,
\begin{align}
	\frac{\partial^2 \ell(\gamma_j)}{\partial \gamma_j^2} = -2 a_j \gamma_j^{-2} + 4b_j\gamma_j^{-3} - \frac{s_j^3\gamma_j+s_j^2-2q_j^2s_j}{\left(\gamma_js_j+1\right)^3},
\end{align}
is negative. If there are multiple real-valued local minima, we choose the one that results in the smallest value of the negative log likelihood $\ell(\gamma_j)$. The final value of the root is denoted $\widetilde{\gamma}_j$.

With this root-finding procedure for determining $\widetilde{\gamma}_j$ in hand, we can now describe the complete FML algorithm. At each iteration, we compute $\widetilde{\gamma}_i$ for $i = 1, \dots, N$. The calculated value of $\widetilde{\gamma}_i$ defines both the \emph{action} we would take to update that element (either \emph{re-estimate}, \emph{add}, or \emph{delete}) and the change in marginal likelihood $\Delta_i$ that performing this action would result in:
\begin{itemize}
	\item If $i \in \mathcal{T}$ and $\widetilde{\gamma}_i > \tau$, \emph{re-est.}: $\gamma_i \leftarrow \widetilde{\gamma}_i$, $\Delta_i = \ell\left(\widetilde{\gamma_i}\right) - \ell\left(\gamma_i\right)$.
	\item If $i \notin \mathcal{T}$ and $\widetilde{\gamma}_i > \tau$, \emph{add}: $\gamma_i \leftarrow \widetilde{\gamma}_i$, $\Delta_i = \ell\left(\widetilde{\gamma_i}\right)$.
	\item If $i \in \mathcal{T}$ and $\widetilde{\gamma}_i \leq \tau$, \emph{delete}: $\gamma_i,~ x_i \leftarrow 0$, $\Delta_i = -\ell\left(\gamma_i\right)$.
\end{itemize}

Once $\Delta_i$ has been calculated for $i = 1, \dots N$, we perform the action that results in the greatest change in marginal likelihood, updating the element at index $j = \arg\max_i \Delta_i$. Performing this action can be performed particularly efficiently because by judiciously representing each element by \cite{tipping2003fast}
\begin{align}
	S_i &= \bm{\phi}_i^T \bm{C}^{-1} \bm{\phi}_i,\quad s_i = \bm{\phi}_i^T \bm{C}_{-i}^{-1} \bm{\phi}_i = \frac{S_i}{1 - \gamma_i S_i} \label{eq:sbl-fml-S-s} \\
	Q_i &= \bm{\phi}_i^T \bm{C}^{-1} \y,\quad q_i = \bm{\phi}_i^T \bm{C}_{-i}^{-1} \y = \frac{Q_i}{1-\gamma_i S_i} \label{eq:sbl-fml-Q-q}
\end{align}
the state updates to $\Sxy$, $\bm{\mu}$, and $\{S_j,Q_j,s_j,q_j\}$ needed to perform the \emph{re-estimate}, \emph{add}, and \emph{delete} actions can be implemented using inexpensive low-rank updates (see Appendix \ref{sec:appendix-fml}). The complete SBL-DF algorithm using the FML method for inference is listed in Algorithm \ref{alg:sbl-df-fml}.

In summary, the FML algorithm initializes an empty model and iteratively re-estimates, adds, or deletes the element that results in the largest improvement of the marginal likelihood. By judiciously representing the state of the model at each iteration using (\ref{eq:sbl-fml-S-s}--\ref{eq:sbl-fml-Q-q}), each update can be implemented using a few inexpensive rank-one updates. In contrast to the FML algorithm of \cite{tipping2003fast}, which is restricted to the uninformative hyperprior SBL probability model, the FML algorithm derived here allows general hyperparamers $a_i$ and $b_i$ to be used at the cost of a more complicated root-finding procedure to compute $\widetilde{\g}$. Because this root-finding procedure must be run for $i = 1, \dots, N$ at each iteration, the total additional cost introduced by the use of informative hyperpriors grows as $O(N)$.

\begin{algorithm}
	\caption{SBL-DF Using Fast Marginal Likelihood}
	\label{alg:sbl-df-fml}
	\begin{algorithmic}[1]
		\State Pick initial index $j= \arg\max_i \abs{\bm{\phi}_i^T \bm{y}^{(1)}}$
		\State Initialize $\Sxy,~\muxy$ using (\ref{eq:sbl-posterior-covariance-mean})
		\State Initialize $S_i, Q_i, s_i, q_i~\forall i$ using (\ref{eq:sbl-fml-S-s}--\ref{eq:sbl-fml-Q-q})
		\For{$t = 1, \dots L$}
			\While{not converged (i.e., $\max_i \Delta_i > \mathrm{tolerance}$)}
				\State Calculate $\left\{\widetilde{\gamma}_i,~\Delta_i\right\}~\forall i$ as in Sec.\ \ref{sec:sbldf-fml}
				\State Pick $j = \arg\max_i \Delta_i$.
				\State Set $\gamma_j$ to $\widetilde{\gamma}_j$ (if \emph{add} or \emph{re-est.}) or $0$ (if \emph{delete})
				\State Update $\Sxy,~\muxy$ using (\ref{eq:sbl-posterior-covariance-mean}).
				\State Update $\{S_i\},~\{Q_i\}$ using Appendix \ref{sec:appendix-fml}.
				\State Update $\{ s_i \}$ and $\{ q_i \}$ using (\ref{eq:sbl-fml-S-s}--\ref{eq:sbl-fml-Q-q})
				\State If desired, update $\lambda$ using (\ref{eq:sbl-em-s2}); recalculate $\Sxy$, $\muxy$, $\{ S_i \}$, and $\{ Q_i \}$
			\EndWhile
			\State Output final posterior mode estimate $\muxy^{(t)}$ as $\widehat{\x}^{(t)}_{\mathrm{SBL}}$
			\State Compute $\bm{b}$ for next time step from $\widehat{\x}^{(t)}$ using (\ref{eq:sbldf-dynamics-mapping})
		\EndFor 
	\end{algorithmic}
\end{algorithm}

\section{Numerical Simulations}
\label{sec:simulations}

In this section, we evaluate the efficacy of the SBL-DF algorithm with synthetic data. These experiments with known ground truth allow us to characterize the performance of SBL-DF in many different regimes.

First, we demonstrate the performance of SBL-DF's method for incorporating dynamics information using a single time step, i.e., we recover a single length-$N$ signal using a noisy signal prediction that mimics the prediction we would obtain using a dynamics model. This allows us to explore SBL-DF's performance as the quality of the dynamics prior information (i.e., accuracy of $\widetilde{\x} = \abs{f_t\left(\widehat{\x}^{(t)}\right)}$) changes. We then show that the single time step performance carries over to tracking problems with multiple time steps.

\subsection{Performance as dynamics prediction quality degrades}
\label{sec:simulations-dynquality}

We first show how incorporating a dynamics-based prediction can improve reconstruction accuracy, and how reconstruction accuracy is affected by the quality of the dynamics-based prediction. In this experiment, we assume that our dynamics-based estimate $\widetilde{\x} = \abs{f_t\left(\widehat{\x}^{(t)}\right)}$ is the ground truth signal corrupted by one of two types of error: (a) support mismatch, in which the values of $\bar{s}$ nonzero elements are swapped with randomly chosen zero-valued elements; and (b) Gaussian noise, in which i.i.d.\ mean-zero Gaussian noise with variance $\sigma_{\mathrm{dyn}}^2$ is added to the entire signal estimate. We generate data using the measurement model (\ref{eq:meas-model}) with $N = 512$, $s = 16$ nonzero elements, $\bm{e} \sim \N\left(\bm{0}, \sigma^2_{\mathrm{obs}}\bm{I}\right)$ with $\sigma^2_{\mathrm{obs}} = 10^{-3}$, and construct the dictionary $\P$ with i.i.d.\ $\N(0,1/\sqrt{M})$ entries. For recovery, we use FML (Algorithm \ref{alg:sbl-df-fml}) with threshold $\tau = 10^{-1}$ and fixed regularization parameter $\lambda = 10^{-3}$. We select the value of the dynamics parameter $\xi$ that minimizes reconstruction error from a fixed grid of $\log_{10}\left(\xi\right)$ from $-2$ to $2$ in increments of $0.1$. As $\xi$ represents how strongly the dynamics-based prediction is considered in the inference procedure, its optimal value is strongly dependent on the quality of the dynamics estimate and therefore a separate value of $\xi$ was selected for each dynamics error level. We vary $M$ for each dynamics noise level, performing $240$ independent trials for each point, and plot the rate of successful recovery (defined as the portion of trials satisfying relative MSE (rMSE) $\norm{\x-\widehat{\x}}_2^2 / \norm{\x}_2^2 < 10^{-2}$).

\begin{figure*}[t]
	\centering
	\begin{tabular}{ccc}
		\includegraphics[height=2.0in]{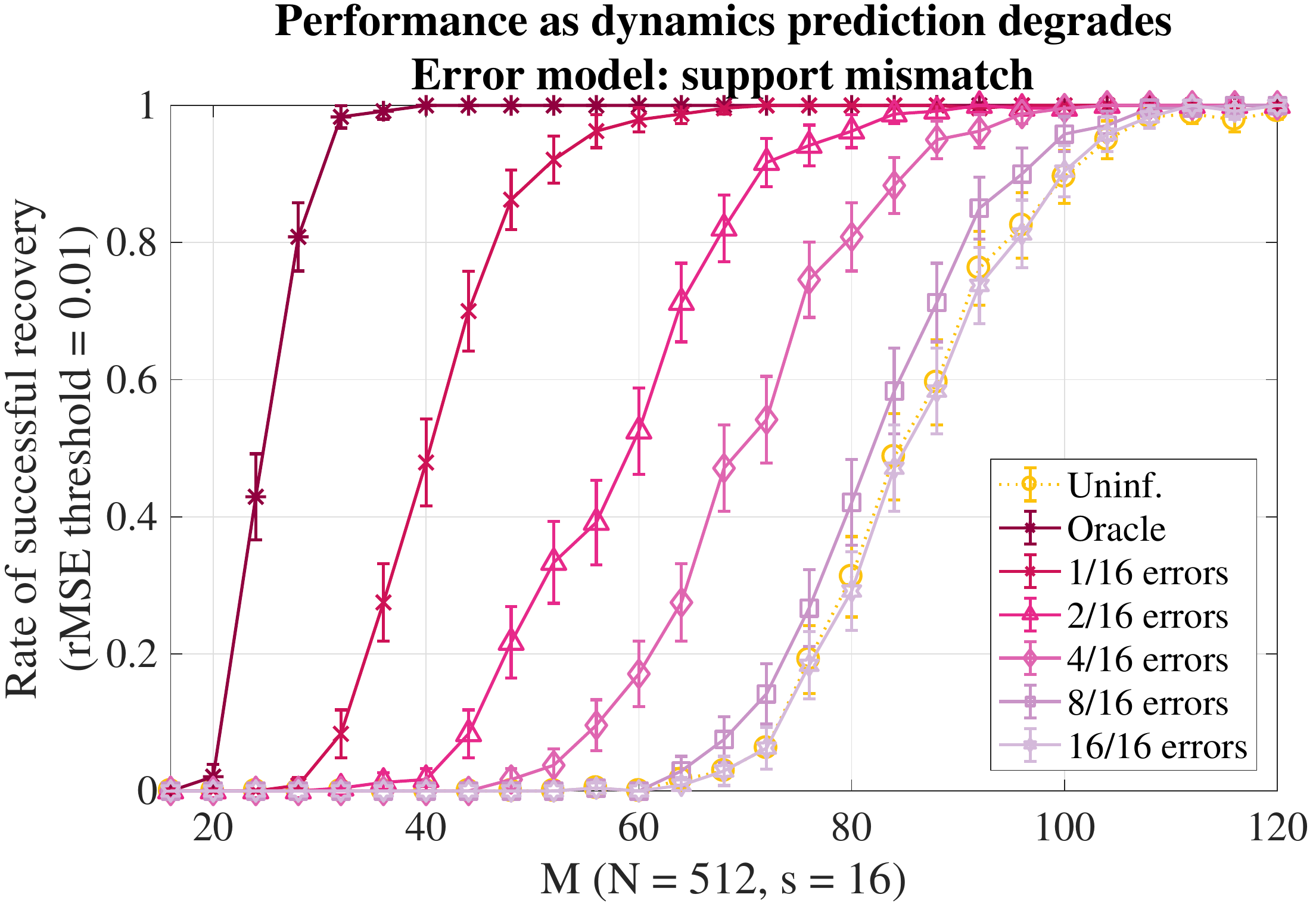} & \hspace{0.0in} &     \includegraphics[height=2.0in]{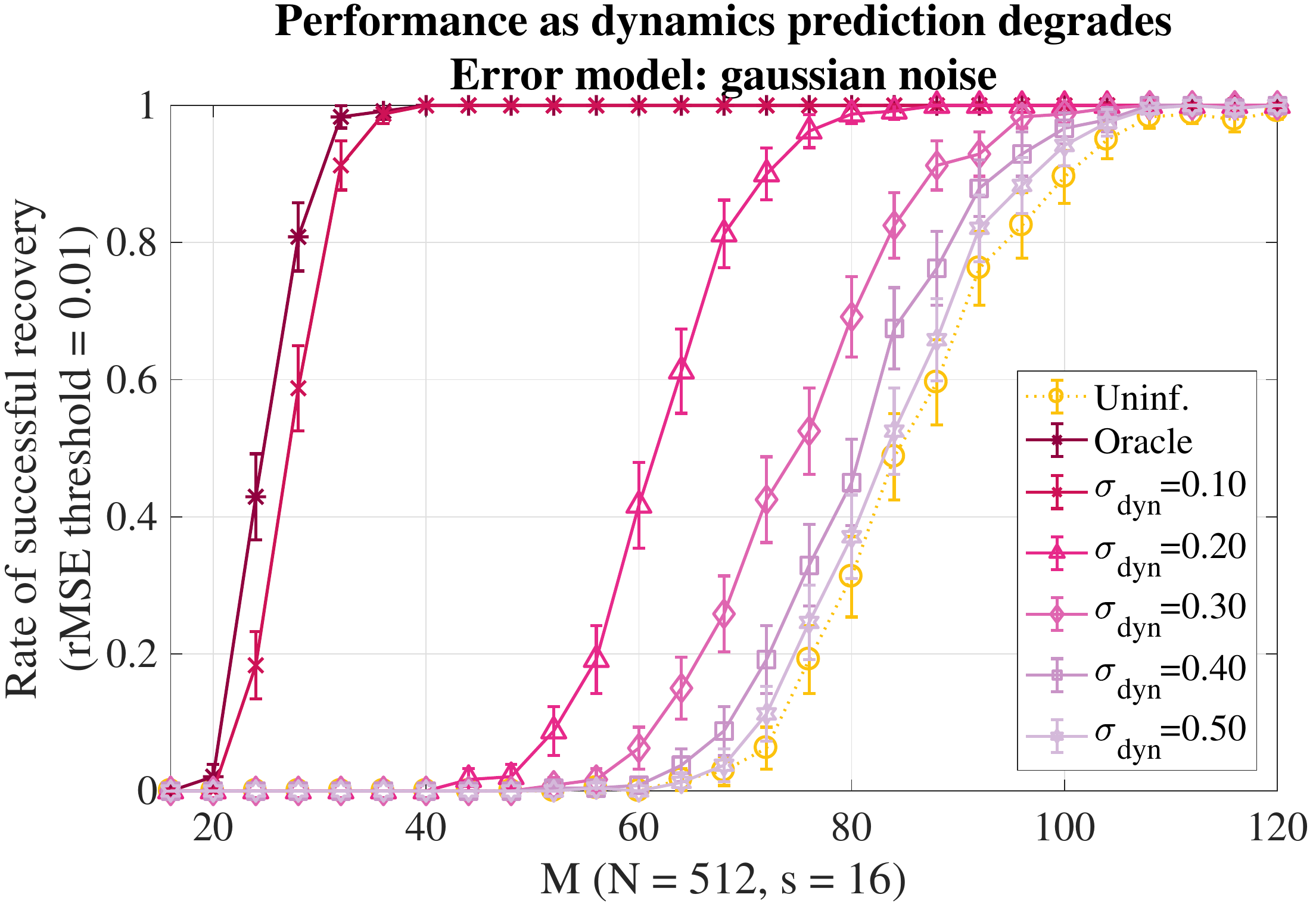}
	\end{tabular}
	\caption{Comparison of rate of successful recovery (defined as relative MSE $\lVert \x - \widehat{\x} \rVert_2^2 / \lVert \x \rVert_2^2 < 0.01$) as a an estimate with increasing amounts of \textit{(Left)} support errors and \textit{(Right)} Gaussian noise is used. Error bars represent a $95\%$ confidence interval on the rate, calculated using the normal approximation. When compared to traditional (uninformative hyperprior) SBL, the number of measurements $M$ required for accurate reconstruction is significantly reduced.}
	\label{fig:singletimestep-accuracy}
\end{figure*}

Figure \ref{fig:singletimestep-accuracy} displays the rate of successful recovery when the prediction $\widetilde{\x}$ is the true signal corrupted with an increasing number of support errors or magnitude of Gaussian noise, respectively. The yellow dotted line represents the performance of the standard uninformative hyperprior SBL algorithm. Both figures demonstrate that SBL-DF's strategy for including a dynamics-based prediction into the SBL probability model results in vastly improved performance when the dynamics-based prediction is accurate in the sense that the true signal can be recovered accurately with far fewer measurements than traditional SBL requires. Further, even when the prediction is highly flawed, SBL-DF with a reasonable choice of $\xi$ still allows $\widehat{\x}^{(t+1)}$ to be accurately recovered with fewer measurements than the standard SBL algorithm.

\begin{figure*}[t]
	\centering
	\begin{tabular}{ccc}
		\includegraphics[height=2.0in]{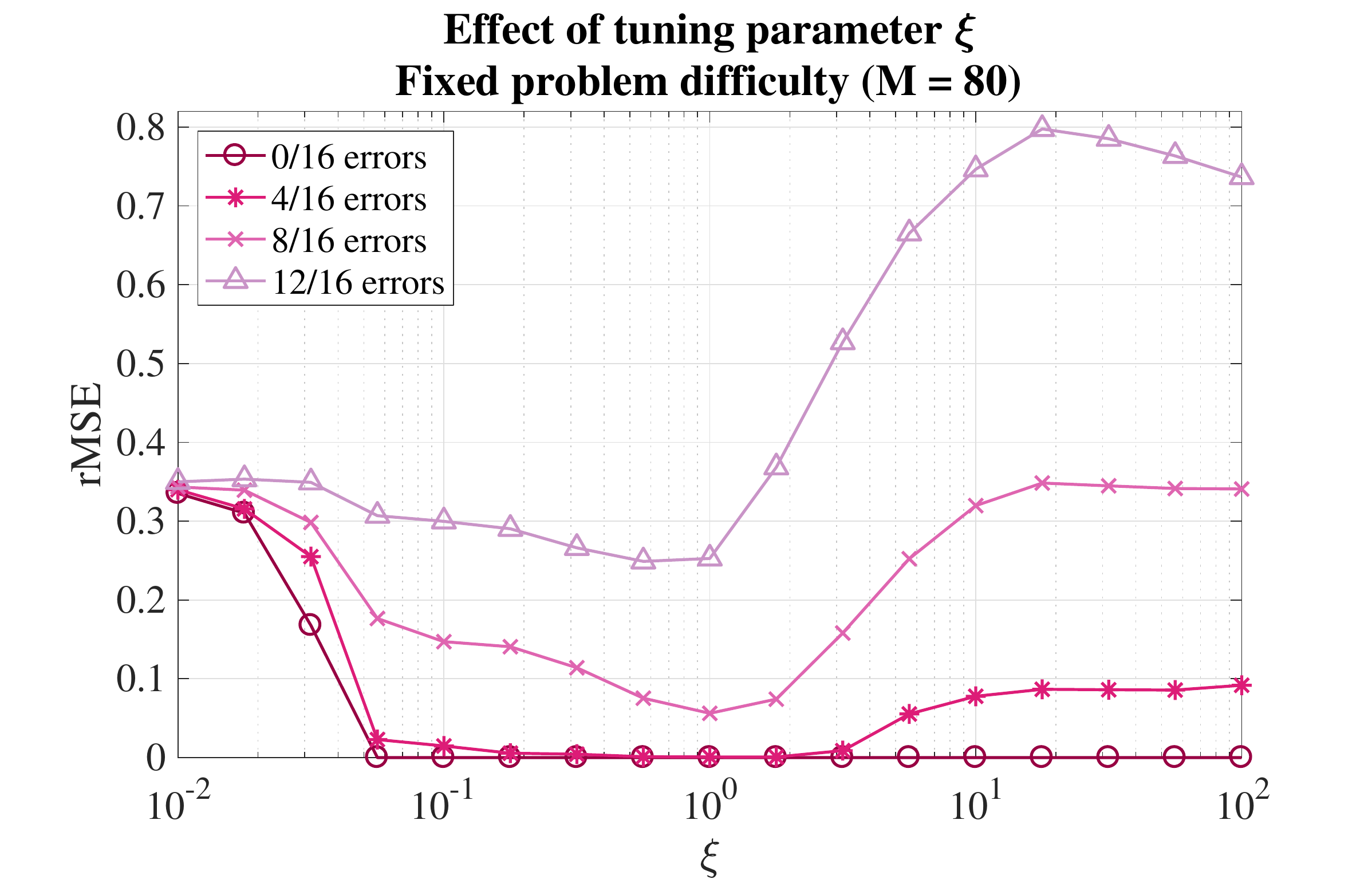} & \hspace{0.0in} &     \hspace{-0.3in}\includegraphics[height=2.0in]{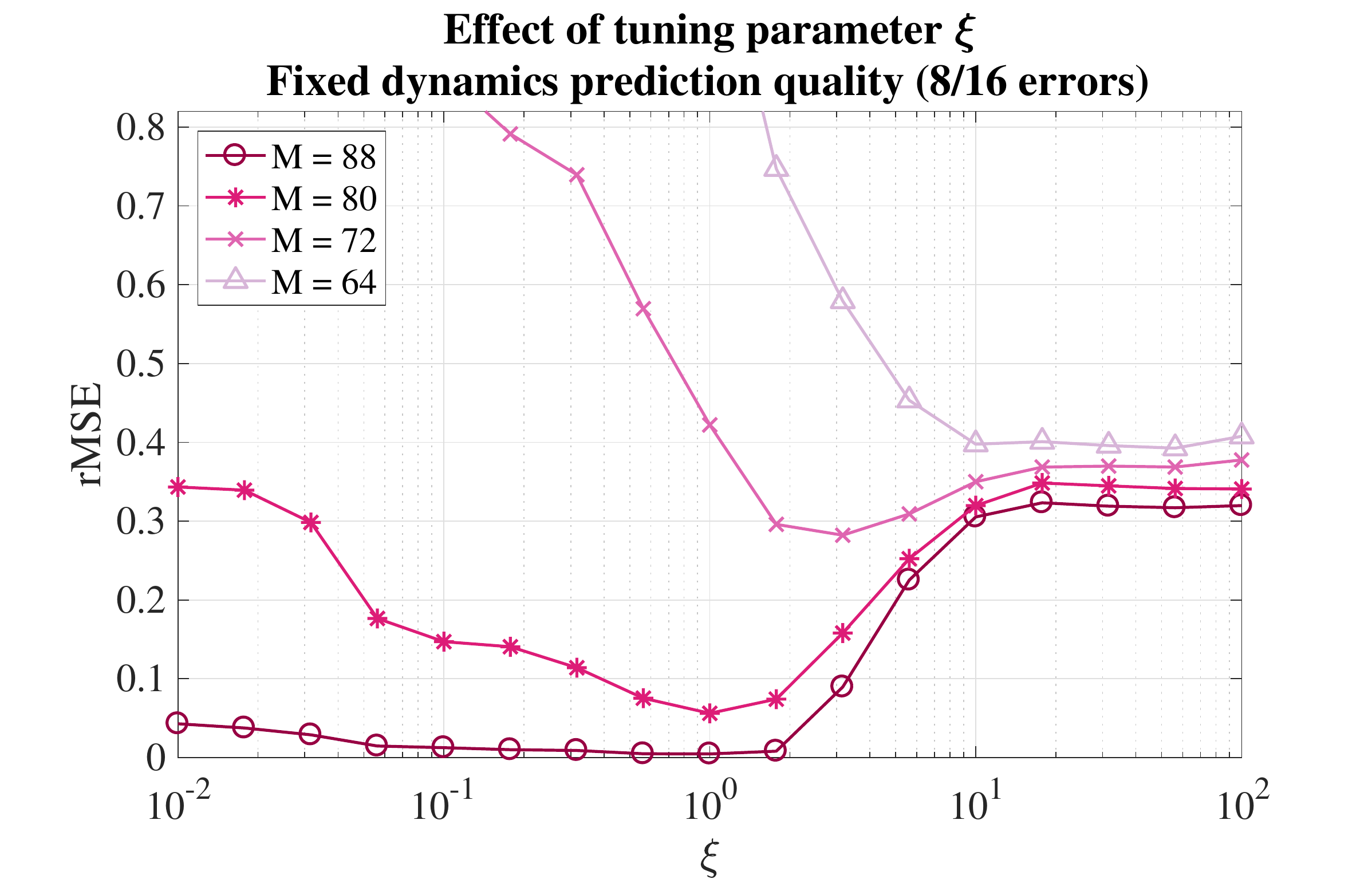}
	\end{tabular}
	\caption{\textit{Left:} Effect of varying the $\xi$ parameter for a fixed problem difficulty. \textit{Right:} Effect of varying the $\xi$ parameter for a fixed dynamics prediction quality. The plots show that the proper choice of $\xi$ depends on both problem difficulty and dynamics model accuracy.}
	\label{fig:singletimestep-sweep-xi}
\end{figure*}

Figure \ref{fig:singletimestep-sweep-xi} examines the effect of changing the $\xi$ parameter (which controls the weight SBL-DF gives the dynamics-based prediction) using the same experimental setup as Figure \ref{fig:singletimestep-accuracy}. In Figure \ref{fig:singletimestep-sweep-xi} (left), we observe that when the dynamics-based prediction $\widetilde{\x}$ is accurate, the reconstruction error (rMSE) decreases monotonically with increasing $\xi$. When the dynamics-based prediction is very inaccurate, there is a value of $\xi$ after which increasing $\xi$ (i.e., enforcing the dynamics-based prediction more strongly in inference) decreases reconstruction accuracy. In Figure \ref{fig:singletimestep-sweep-xi} (right), we observe that when the dynamics-based prediction $\widetilde{\x}$ is accurate, the reconstruction error (rMSE) decreases monotonically with increasing $\xi$. However, when the dynamics-based prediction is noisy, there is a value of $\xi$ after which increasing $\xi$ decreases reconstruction accuracy.

\subsection{Performance with structured dictionaries}

Next, we explore the performance of SBL-DF when the dictionary contains challenging structure such as highly correlated columns or column norms of drastically different magnitudes. Although this type of dictionary structure is common in many practical applications (e.g., neural imaging \cite{wipf2011sparse}), $\ell_1$ minimization-based algorithms provably fail in this setting. In contrast, traditional SBL has been shown to be provably and empirically robust to this type of dictionary structure, and in fact SBL is commonly used in applications with unfavorably structured dictionaries \cite{wipf2011sparse,zhang2011clarify,he2017bayesian}. In this experiment, we explore how this robustness extends to the dynamic case where prior knowledge is incorporated. Specifically, we compare the SBL-DF algorithm with RWL1-DF \cite{charles2016dynamic}, a state of the art $\ell_1$ minimization based tracking algorithm based on reweighted $\ell_1$ \cite{candes2008enhancing} when this type of undesirable dictionary structure is present.

We consider four dictionary models constructed from $\widetilde{\P} \in \R^{M \times N}$ containing i.i.d.\ $\N(0,1/\sqrt{M})$ entries:
\begin{itemize}
	\item \emph{i.i.d.}: $\P = \widetilde{\P}$.
	\item \emph{i.i.d.\ with scaled columns}: $\P = \widetilde{\P} \bm{D}$, where $\bm{D} \in \R^{N \times N}$ is the diagonal matrix with i.i.d.\ $U[0,1]$ entries on the diagonal.
	\item \emph{Locally coherent}: $\P = \widetilde{\P} \bm{B}$, where $\bm{B} \in \R^{N \times N}$ is a block diagonal matrix consisting of $4 \times 4$ blocks, each with $1$ on its diagonal and $0.8$ elsewhere. This model, which is similar to the setup of the structured dictionary in \cite{he2017bayesian}, results in a dictionary with $25$ groups of $4$ highly correlated columns.
	\item \emph{Locally coherent with scaled columns}: $\P = \widetilde{\P} \bm{B} \bm{D}$, where $\bm{B}$ and $\bm{D}$ are defined as above. This model matches the setup of the structured dictionary in \cite{he2017bayesian}.
\end{itemize}
To standardize the effective SNR of each model, we scale each of the resulting dictionaries by the scalar $\lVert\widetilde{\P}\x\rVert_2 / \lVert\P\x\rVert_2$.

We also consider two models for the signal $\x$. First, we consider the model where the $s$ nonzero elements of $\x$ are i.i.d.\ $\N(0,1)$. This setting, in which SBL has been shown to outperform other sparse estimation methods \cite{wipf2011latent}, is frequently considered in the SBL literature. Second, we consider the model where the $s$ nonzero elements of $\x$ all have value $1$. This setting is more frequently considered in traditional compressive sensing literature.

In this experiment, we set $N = 100$, $M = 42$, $s = 25$, and sweep the observation noise $\sigma^2_{\mathrm{obs}}$ from $10^{-8}$ to $10^{-3}$. Note that although this is a relatively small amount of noise in the prototypical i.i.d.\ dictionary setting, the structured dictionary models we consider here make reconstruction extremely challenging at this noise level. To simulate the dynamics-based prediction $\widetilde{\x}$, we add support errors to $\x$ by switching the value of each nonzero element of $\x$ with a zero-valued element with probability $p = 0.1$ and then add white Gaussian noise with variance $\sigma_{\mathrm{dyn}}^2 = 10^{-4}$ to every element of $\widetilde{\x}$. We perform SBL-DF inference using the EM updates of Algorithm \ref{alg:sbl-df-em} with pruning threshold $\tau = 10^{-4}$, select the dynamics parameter $\xi$ by a grid search run independently for each point, and use the automatic learning rule (\ref{eq:sbl-em-s2}) to select the regularization parameter $\lambda$. For RWL1-DF, the three parameters $\lambda$, $\xi$, $\eta$ were selected by a grid search run independently for each point.

\begin{figure*}[t]
	\centering
	\includegraphics[width=\textwidth]{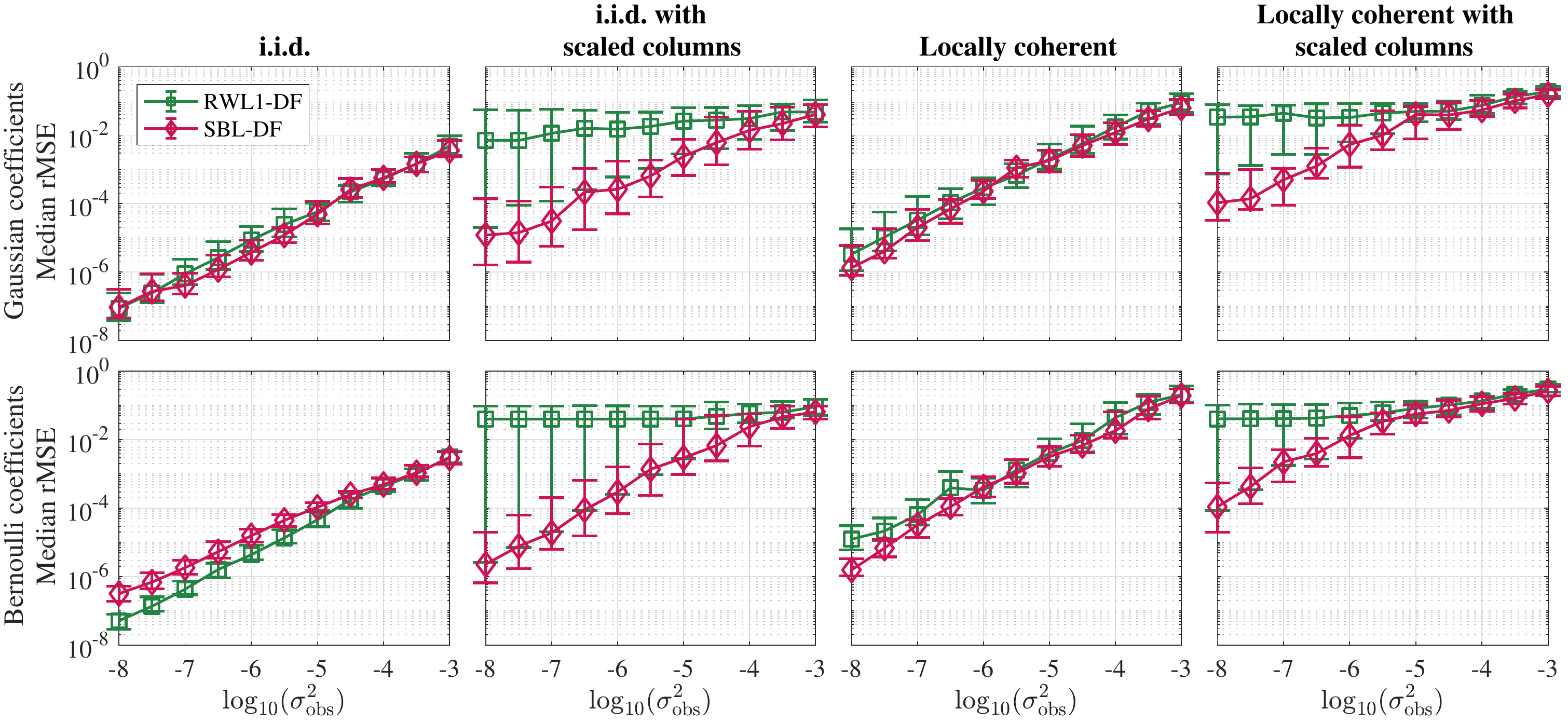}
	\caption{Reconstruction accuracy for SBL-DF and RWL1-DF as the observation noise increases for four dictionary models. \emph{Top}: nonzero coefficients in $\x$ drawn from $\N(0,1)$. \emph{Bottom}: nonzero coefficients in $\x$ all unity-valued. Main line shows median rMSE of $20$ independent trials; error bars show $25$th and $75$th percentile rMSE. SBL-DF outperforms the state-of-the-art RWL1-DF algorithm when using the i.i.d.\ with scaled columns and locally coherent with scaled columns dictionary models. With the locally coherent dictionary model, SBL-DF has approximately the same median error as RWL1-DF, but with significantly lower variance.}
	\label{fig:singletimestep-coherence-sweeps2}
\end{figure*}

Figure \ref{fig:singletimestep-coherence-sweeps2} shows the rMSE $\norm{\x-\widehat{\x}}_2^2/\norm{\x}_2^2$ as the observation noise variance $\sigma_{\mathrm{obs}}^2$ increases. In this figure, the main line shows the median rMSE, and error bars show the $25$th and $75$th percentile rMSE to give a sense of the reconstruction error variance. We observe that with an i.i.d.\ dictionary, both RWL1-DF and SBL-DF are able to reconstruct the signal accurately, with error beginning to increase to unacceptable levels only as $\sigma_{\mathrm{obs}}^2$ grows large. However, when the dictionary has the challenging types of structure considered here, SBL-DF tends to perform better than RWL1-DF, recovering the signal with lower error (for i.i.d.\ with scaled columns and locally coherent with scaled columns models) or lower variance (for the locally coherent model) at high to moderate SNRs. We have observed that SBL-DF similarly outperforms RWL1-DF when using other models of coherent dictionaries, such as dictionaries containing i.i.d.\ Gaussian entries with nonzero mean. The robustness displayed by SBL-DF aligns with theoretical results in the static SBL literature, which has shown that in the absence of measurement noise, the uninformative hyperprior SBL objective is invariant to column scaling and coherence structure. In contrast, the objective of $\ell_1$ based methods are invariant to either column scaling or coherence only in isolation \cite{wipf2011sparse}.

In addition to superior reconstruction accuracy, SBL-DF also has the significant advantage of only requiring the tuning of a single parameter\footnote{Although the pruning threshold $\tau$ could nominally be considered a tuning parameter, it can be typically be set to any reasonable value without a large change in results. Here, we fix $\tau = 10^{-4}$ in all experiments using the EM algorithm.}. In contrast, RWL1-DF requires tuning three parameters. Since in this experiment we perform a grid search over only one parameter for SBL-DF, and a simultaneous grid search over three parameters for RWL1-DF, the results in this experiment are charitable for RWL1-DF and some degradation in RWL1-DF's performance would be expected in applications where it is not possible to perform this large grid search.

\begin{figure}[t]
	\centering
	\includegraphics[width=0.6\textwidth]{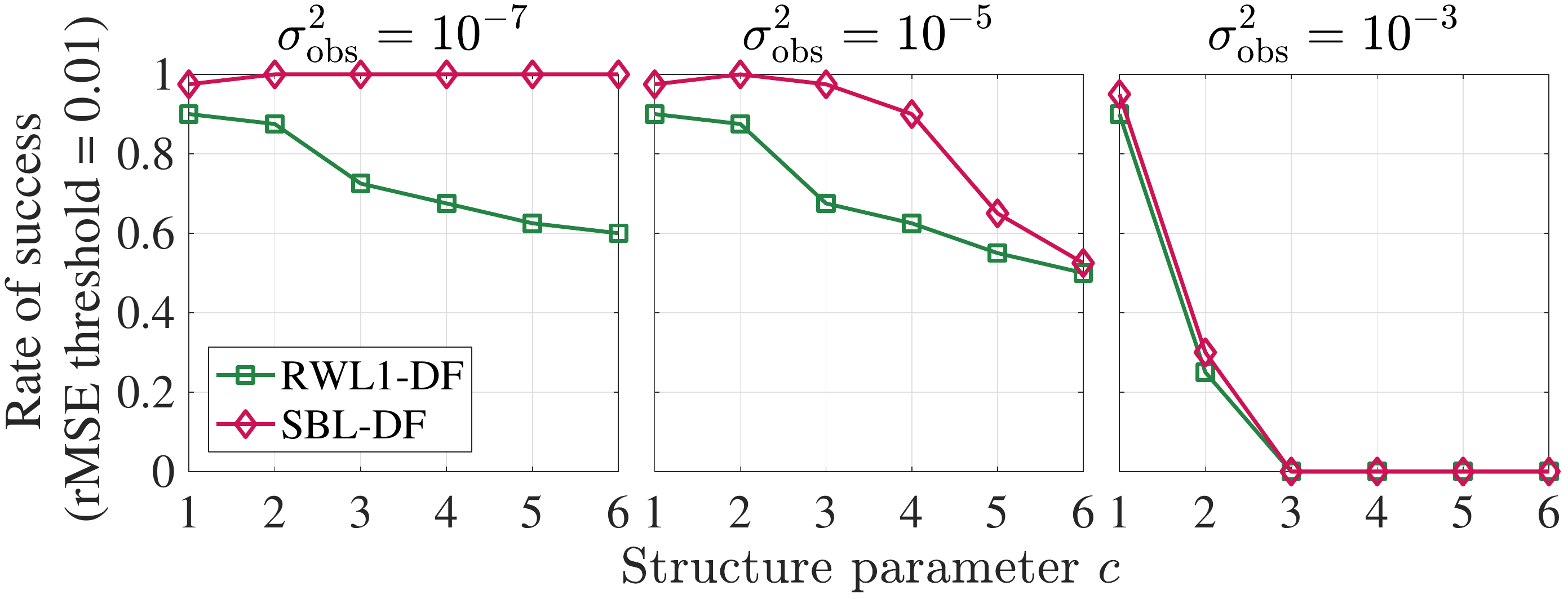}
	\caption{Reconstruction accuracy for SBL-DF and RWL1-DF as the dictionary structure parameter (which controls dictionary coherence and column scaling; see text for details) increases. SBL-DF outperforms existing algorithms on these challenging problems with structured dictionaries and small amounts of observation noise.}
	\label{fig:singletimestep-coherence-sweepcoh}
\end{figure}

We further explore SBL-DF's performance in the presence of challenging dictionary structure in Figure \ref{fig:singletimestep-coherence-sweepcoh}. In this experiment, we again set $N = 100$, $M = 42$, $s = 25$, but now fix $\sigma_{\mathrm{obs}}^2$ to three different values. As in the previous experiment, the dynamics-based prediction is the ground truth signal corrupted with support errors occurring with probability $p = 0.1$ and white Gaussian noise with $\sigma_{\mathrm{dyn}}^2 = 10^{-4}$. The dictionary is constructed using the ``locally coherent with scaled columns'' model described above, but we now sweep the amount of structure in the dictionary. Specifically, each block of $\bm{B}$ now takes value $1-\tfrac{1}{c}$ off its diagonal and the entries of $\bm{D}$ are i.i.d.\ $U[\tfrac{1}{c},1]$ so that a larger value of $c$ results in a larger range of column scales. Recall that $\P$ is scaled to have the same norm as $\widetilde{\P}$, so the overall scaling of $\D$ does not affect the SNR. This parameterization results in the dictionary being equivalent to the i.i.d.\ model described above when $c = 1$; as $c$ increases, the dictionary becomes increasingly structured. For each noise level, we sweep the structure parameter $c$, performing $40$ independent trials at each point and plotting the rate of successful recovery (defined as rMSE $\norm{\x-\widehat{\x}}_2^2 / \norm{\x}_2^2 < 10^{-2}$). As before we sweep the single parameter $\xi$ for SBL-DF using a grid search for each noise and structure level, and sweep the three parameters, $\lambda$, $\xi$, and $\eta$, for RWL1-DF using a simultaneous grid search for each noise and structure level.

From Figure \ref{fig:singletimestep-coherence-sweepcoh} we observe that, in general, increasing dictionary structure makes it more difficult for both algorithms to accurately recover the signal. However, in contrast to RWL1-DF, in high SNR regimes SBL-DF is able to accurately recover the signal even with very highly structured dictionaries. As the SNR becomes lower and reconstruction becomes increasingly difficult with any amount of structure, both algorithms are able to accurately recover the signal when the dictionary has i.i.d.\ entries ($c = 1$), but not when the amount of structure increases.

\subsection{Runtime comparison}

Next, we compare the runtime of SBL and SBL-DF using both the EM and FML inference procedures. In this experiment we fix the underdeterminedness ratio to $M/N = 1/4$ and the number of nonzero entries to $s = 16$, and sweep the signal length from $N = 2^{9}$ to $N = 2^{14}$. We use the dictionary $\P \in \R^{M \times N}$ containing i.i.d.\ $\N(0,1/\sqrt{M})$ entries, set the observation noise variance to $\sigma^2_{\mathrm{obs}} = 10^{-3}$, and generate the dynamics-based prediction as the true signal corrupted with support errors occurring with probability $p = 0.1$ and Gaussian noise with variance $\sigma_{\mathrm{dyn}}^2 = 10^{-4}$. As in the previous experiments, we set the pruning threshold to $\tau = 10^{-4}$ for EM and $\tau = 10^{-1}$ for FML, fix $\lambda = 1.2 \times 10^{-3}$ for both EM and FML, and set $\xi = 10^{0}$ for SBL-DF. Each algorithm is run until convergence\footnote{The FML algorithm maintains and updates values that can be used to efficiently calculate the change in the log likelihood at each iteration, which is used to determine convergence. In contrast, our EM implementation uses $\norm{\muxy^{\mathrm{(new)}} - \muxy}_2$ as the convergence criteria as it does not maintain the value of the log likelihood. Therefore, to facilitate a fair comparison, in these experiments we use $\norm{\muxy^{\mathrm{(new)}} - \muxy}_2 < \mathrm{tol}$ as the convergence criteria for both algorithms.} with a tolerance of $10^{-4}$. This tolerance was selected as it is approximately the value where the algorithm achieves its final error. For this experiment, SBL (EM) and SBL (FML) had similar rMSE, and SBL-DF (EM) and SBL-DF (FML) had similar rMSE. Each trial in this section was run on a single core of an Intel Xeon E5-2680 v3 processor with a base clock rate of 2.5GHz and a shared pool of 100GB memory.

\begin{figure}[t]
	\centering
	\includegraphics[width=0.6\textwidth]{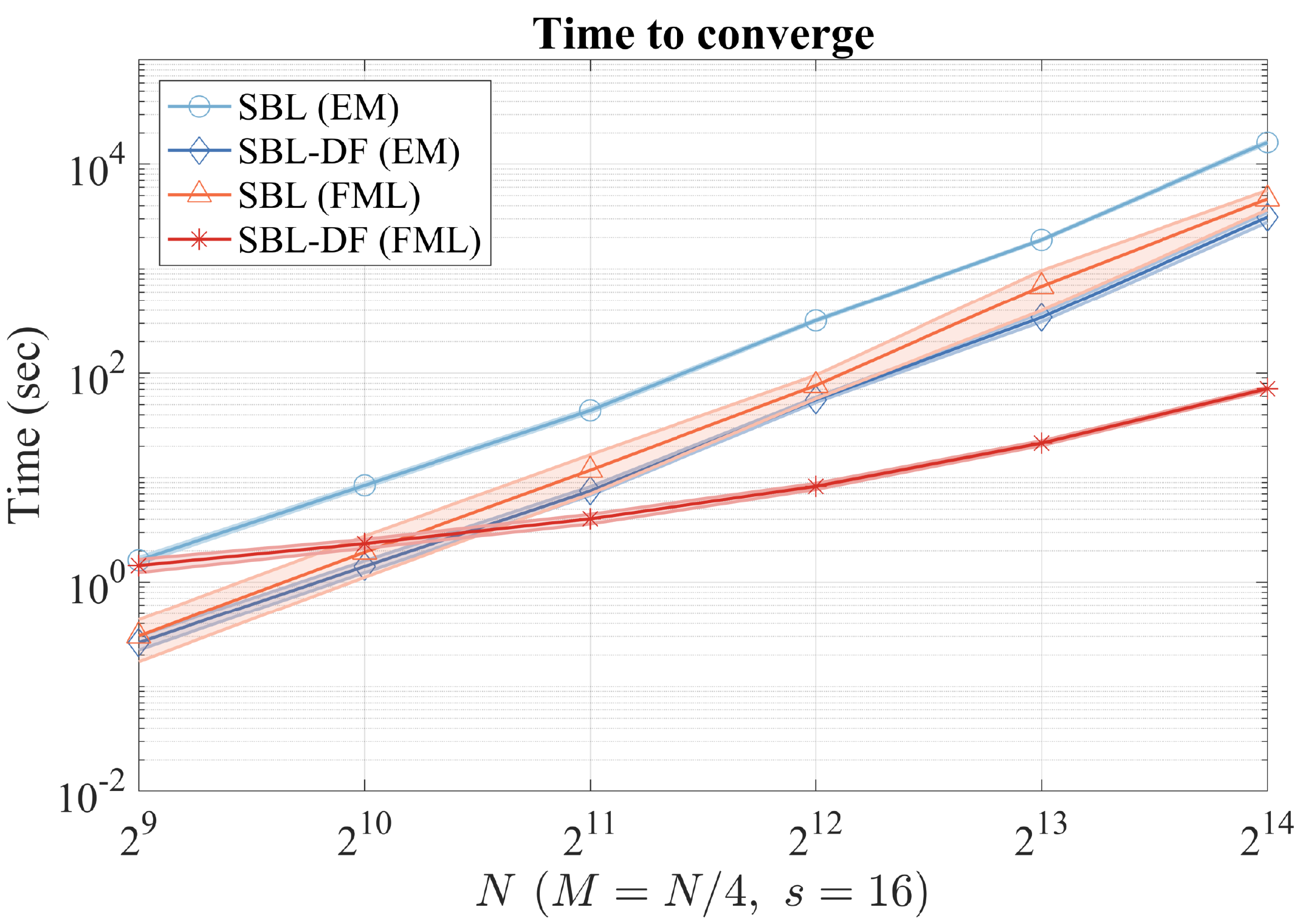}
	\caption{Reconstruction time for the SBL and SBL-DF algorithms using the EM and FML method as the problem size ($M$ and $N$) increases. Each point displays the mean and standard deviation of $24$ independent trials.}
	\label{fig:singletimestep-time}
\end{figure}

\begin{figure}[t]
	\centering
	\includegraphics[width=0.6\textwidth]{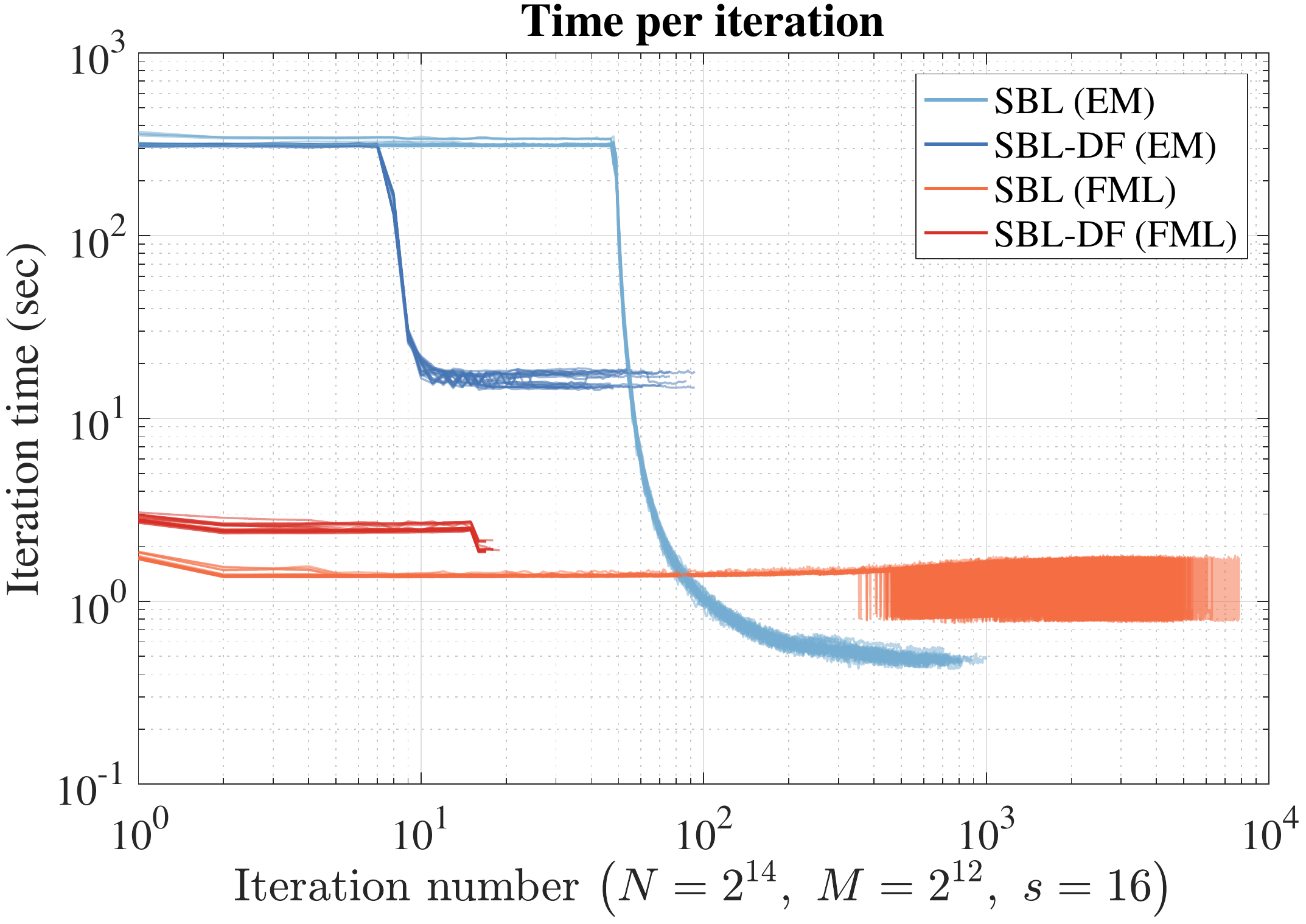}
	\caption{Reconstruction time as the number of nonzero entries increases. Each trace is the result of one independent trial.}
	\label{fig:singletimestep-time-iteration}
\end{figure}

Figure \ref{fig:singletimestep-time} shows the total execution time for each algorithm as the problem size increases, with each point displaying the mean of $24$ independent trials. We first compare the execution times of SBL (EM) and SBL-DF (EM). We see that despite the fact that the initial EM iterations for SBL and SBL-DF have identical $O\left(\widetilde{N}^3 + \widetilde{N}^2M + \widetilde{N}^2 + \widetilde{N}M + \widetilde{N}\right)$ per-iteration cost, SBL-DF's incorporation of dynamics information with informative hyperpriors reduces the number of iterations required for convergence by approximately one order of magnitude. Figure \ref{fig:singletimestep-time-iteration}, which shows the execution time of each iteration, suggests that this improvement is due to SBL-DF inferring and pruning zero-valued elements (and therefore reducing $\widetilde{N}$) much more quickly than SBL.

The relative performance of SBL (FML) and SBL-DF (FML) is slightly more nuanced. The incorporation of informative hyperpriors into the FML method for SBL-DF requires a root-finding procedure that adds a performance penalty to each iteration of SBL-DF (FML) when compared to SBL (FML). However, the cost of this root-finding procedure grows only linearly with $N$, and as the problem size grows this cost is dominated by the $O\left(\widetilde{N}^2+MN+N\right)$ matrix operations required for computing the low-rank state updates in each iteration. Additionally, Figure \ref{fig:singletimestep-time-iteration} shows that the incorporation of prior information allows SBL-DF to converge with many fewer iterations of the ``re-estimate'' action. The overall effect of these two phenomena is that SBL-DF (FML) is more expensive than SBL (FML) on small problems, but becomes more efficient than SBL (FML) as $N$ and $M$ increase.

Comparing the execution times of the EM and FML inference procedures, we observe that when using standard (uninformative hyperprior) SBL, the FML method is both significantly more efficient than EM and scales slightly better with the problem size. When using informative hyperpriors (as in SBL-DF), the FML method requires an additional computational burden from the added complexity of root calculation, but has execution time that scales significantly better with problem size than EM. We have observed that, unlike the EM algorithm, the number of iterations required by the FML method is strongly dependent on the sparsity level, so the EM method may still be preferable in large problems where $s/N$ is relatively large. Further, the convergence speed of SBL-DF is dependent on the quality of the dynamics estimate, with SBL-DF converging more quickly when the dynamics estimate is close to the true signal.

\subsection{Dynamic filtering}

Here, we demonstrate the efficacy of SBL-DF on an $L=30$ time step synthetic tracking problem. The following procedure (similar to that used in \cite{charles2016dynamic}) is used to generate synthetic tracking data. First, $\x^{(1)} \in \R^N$ is generated with $s$ nonzero targets drawn from $\N\left(0,1\right)$, but with elements where $\abs{x_i} < 0.1$ set to $0.1\times\mathrm{sign}(x_i)$. To generate the remaining time steps, $\x^{(2)}, \dots, \x^{(L)}$, each target is assigned a direction (i.e., $-1$ or $+1$), and the linear dynamics model $\bm{F}$ is generated to correspond to each target moving one index in its assigned direction at each time step. However, the actual signal experiences sparse innovations, with each element moving in the opposite direction with probability $p = 0.1$ at each time step. We set $N = 100$, $M = 42$, $s = 25$, $\sigma^2_{\mathrm{obs}} = 10^{-6}$, and generate the dictionary using the ``locally coherent with scaled columns'' model described above. We compare to static SBL, static RWL1 \cite{candes2008enhancing} and RWL1-DF \cite{charles2016dynamic}. In addition, we compare to the windowed TMSBL (WTMSBL) algorithm \cite{zhang2011exploiting}, which runs the TMSBL multiple measurement vector SBL algorithm on a series of overlapping windows and is representative of a body of literature that assumes the true signal is approximately stationary over small intervals.

\begin{figure}[t]
	\centering
	\includegraphics[width=0.6\textwidth]{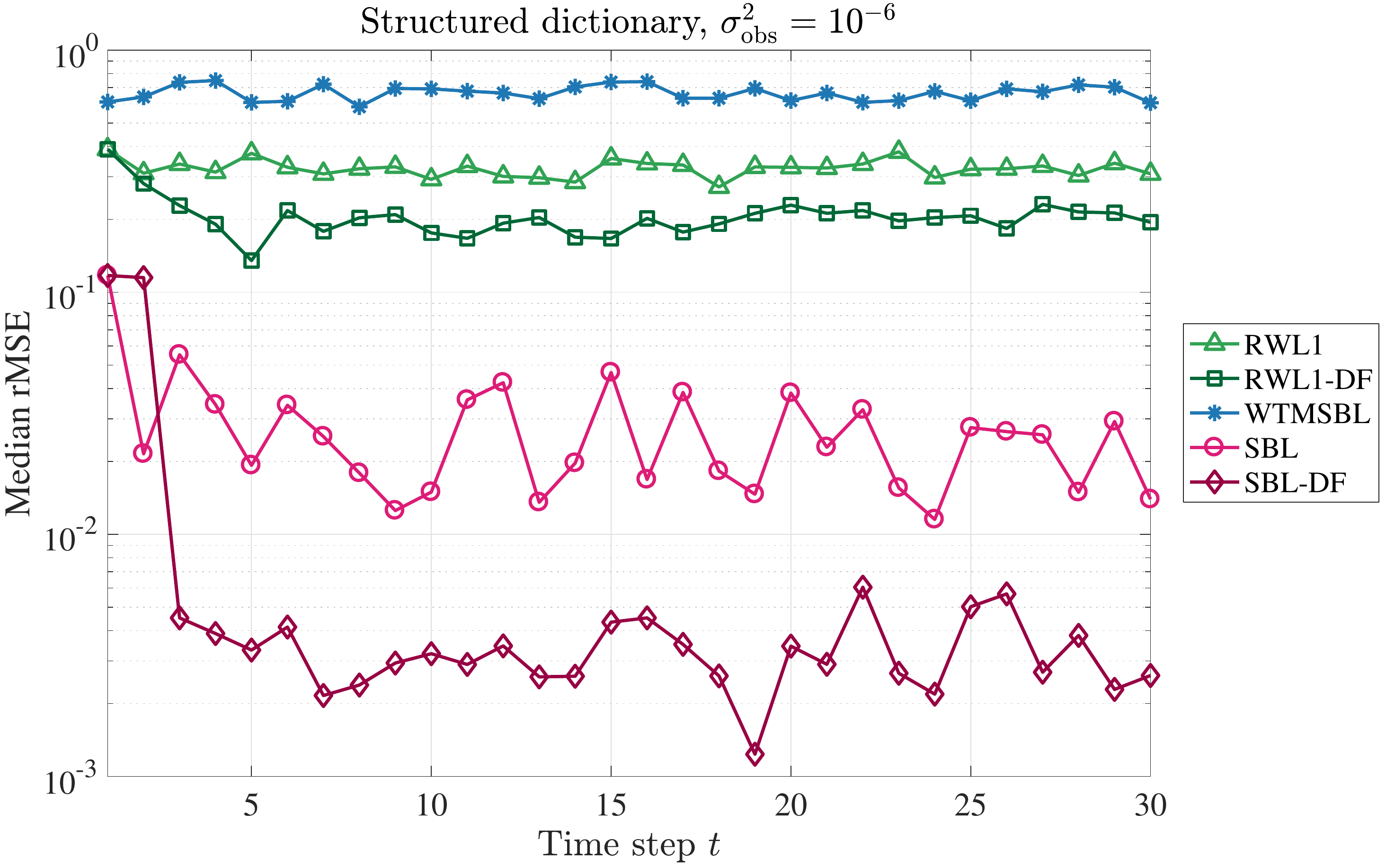}
	\caption{Relative error of each algorithm at each time step in a $L=30$ time step synthetic tracking example. SBL-DF's incorporation of prior knowledge and a noisy dynamics model results in a lower MSE than SBL performed at each time step. In this problem, the windowed TMSBL (WTMSBL) approach of \cite{zhang2011exploiting} does not perform well because the signal violates the slowly time varying model.}
	\label{fig:tracking-time}
\end{figure}

Figure \ref{fig:tracking-time} shows the results of the multiple time step dynamic filtering simulation. We observe that both RWL1-DF and SBL-DF outperform their static counterparts that do not take advantage of a dynamics based signal estimate. However, SBL literature has shown that this type of dictionary structure is particularly detrimental to $\ell_1$-based algorithms, and here we observe that static SBL outperforms both static RWL1 and RWL1-DF. SBL-DF further improves the performance of SBL, allowing accurate reconstruction even in the presence of challenging dictionary structure and observation noise.  We observe that, due to the challenging structure in the dictionary \cite{wipf2011sparse}, static SBL outperforms RWL1. The windowed TMSBL method performs worse than the standard single measurement vector methods in this setting because the signal violates the slowly time-varying assumption. Both RWL1-DF and SBL-DF are able to improve reconstruction performance over their static counterparts in this setting by using an (imperfect) dynamics model.



\section{Discussion}
\label{sec:discussion}

In this paper, we proposed an algorithm for tracking time-varying sparse signals using the sparse Bayesian learning (SBL) framework. We demonstrated that our method for incorporating prior state estimates and an imperfect dynamics model into the SBL probability model allows the algorithm to recover signals more accurately and efficiently than previous methods. Further, we showed that the algorithm outperforms state of the art $\ell_1$-based methods when the dictionary contains challenging structure such as coherence and diverse column magnitudes. In addition, we adapted the fast marginal likelihood \cite{tipping2003fast} and reweighted $\ell_1$ \cite{wipf2010iterative} procedures to the informative hyperprior SBL setting, allowing their use with SBL-DF and allowing us to draw comparisons to static SBL.

In this work, we applied our proposed method to the causal reconstruction of time-varying signals, where the signal estimate is obtained from previously recovered time steps and a dynamics model. However, our general framework for mapping a signal estimate into the SBL probability model developed in Section \ref{sec:sbldf} is applicable to any problem where an a priori estimate of a signal can be used to improve inference.

There are several avenues for future work. One potential improvement could be made by incorporating related work estimating correlation structure or dynamics models from data, allowing the algorithm to benefit from trends automatically discovered in previous time steps. SBL-DF is particularly well-poised to take advantage of research advances in this area due to its use of an arbitrary dynamics model. Additionally, SBL-DF's probabilistic formulation allows other priors or signal models to be easily incorporated into the algorithm; extensions could add application-specific domain knowledge or signal characteristics such as cluster structure.

A limitation of SBL-DF is inherited from the lack of strong theoretical performance and convergence guarantees in SBL literature, which are not currently as developed as those for $\ell_1$-based methods. Theoretical advances in this direction could lead to improved understanding of where and why SBL-DF outperforms similar $\ell_1$-based methods, and more rigorously quantify SBL-DF's performance relative to static SBL and other dynamic filtering algorithms.

Finally, while we have provided a fast algorithm based on the widely used fast marginal likelihood updates, leveraging recent work performing on SBL inference using approximate message passing \cite{al-shoukairi2018gampbased} or variational methods \cite{karseras2014fast} has the potential to admit implementations of SBL-DF that are highly efficient in high-dimensional settings or more amenable to modifications of the SBL probability model.

\bibliographystyle{IEEEtran}
\bibliography{IEEEabrv,moshaughnessy-refs.bib}

\appendix 

\section{Derivation of Reweighted $\ell_1$ Implementation}
\label{sec:appendix-rwl1-sbl}

As in \cite{wipf2008new,wipf2010iterative}, we take a majorization-minimization approach to minimizing (\ref{eq:sbl-objective}). We begin by constructing a function that majorizes the SBL cost function (\ref{eq:sbl-objective}) as a sum of two upper bounding functions: one that bounds the concave expression $\log \abs{\Sy} - 2 \sum_i a_i \log \gamma_i^{-1}$, and one that bounds the convex expression $\y^T \Sy^{-1} \y + 2 \sum_i b_i \gamma_i^{-1}$.

First, we bound $\log \abs{\Sy} - 2\sum_ia_i\log\gamma_i^{-1}$ as
\begin{equation}
\log \abs{\Sy} - 2 \sum_i a_i \log \gamma_i^{-1} \leq \z^T\g - g^*(\z), \label{eq:sbldf-rwl1-bound-concave}
\end{equation}
where $g^*(\z) = \inf_{\g} \left[ \z^T\g - g(\g) \right]$ is the \emph{concave conjugate} of $g(\g) = \log \abs{\Sy} - 2 \sum_i a_i \log \gamma_i^{-1}$ (see \cite{wipf2010iterative} for details). Intuitively, $\z^T\g - g^*(\z)$ is a family of linear upper bounds on $g(\g)$; the tightest bound at a fixed value of $\g$ is achieved by minimizing over $\z$. Note that we do not need to explicitly calculate $g^*(\z)$ because we will only need to minimize the right hand side of (\ref{eq:sbldf-rwl1-bound-concave}) with respect to $\g$.

Second, we bound the convex expression $\y^T\Sy^{-1}\y + 2\sum_ib_i\gamma_i^{-1}$. By applying the matrix inversion lemma and recognizing the form of the result as the solution to the of a Tikhonov-regularized least-squares problem, $\y^T \Sy^{-1} \y$ can be written as \cite{wipf2010iterative} $\y^T\Sy^{-1}\y = \underset{\x}{\arg\min}~\frac{1}{\lambda} \norm{\y-\P\x}_2^2 + \norm{\G^{-1/2}\x}_2^2$, giving the bound
\begin{align}
	&\y^T\Sy^{-1}\y + 2\sum_ib_i\gamma_i^{-1} \notag \\
	&\qquad\leq \frac{1}{\lambda} \norm{\y-\P\x}_2^2 + \norm{\G^{-1/2}\x}_2^2 + 2\sum_ib_i\gamma_i^{-1}. \label{eq:sbldf-rwl1-bound-convex}
\end{align}

Combining (\ref{eq:sbldf-rwl1-bound-concave}) and (\ref{eq:sbldf-rwl1-bound-convex}) gives the majorizing function\footnote{We have suppressed the dependence on $\lambda$ in $\ell_{\z}\left(\g,\x\right)$ and $\ell\left(\g\right)$ for clarity.}
\begin{align}
	\ell_{\z}\left(\g,\x\right) &= \z^T\g - g^*(\z) + \frac{1}{\lambda} \norm{\y-\P\x}_2^2 \notag \\
	&\qquad + \x^T\G^{-1}\x + 2\sum_ib_i\gamma_i^{-1} \geq \ell(\g). \label{eq:sbldf-rwl1-majorizer}
\end{align}

We now construct the majorization-minimization iterations. The majorization step updates the majorizing function by calculating the $\z$ that results in the tightest bound (\ref{eq:sbldf-rwl1-majorizer}) at $\g$. From duality theory, this value of $\z$ occurs at the gradient of $g(\z)$ \cite{jordan1999introduction}, so the majorization step is
\begin{align}
	z^{(k+1)}_i &= \frac{\partial}{\partial \gamma_i} \left[ \log \abs{\Sy} - 2 \sum_i a_i \log \gamma_i^{-1} \right] \notag \\
	&= \bm{\phi}_i^T\Sy^{-1}\bm{\phi}_i + \frac{2 a_i}{\gamma_i^{(k)}}. \label{eq:sbldf-rwl1-z}
\end{align}

The minimization step consists of solving
\begin{align}
	\x^{(k+1)} &= \underset{\g,\x}{\arg\min}~\ell_{\z}\left(\g,\x\right) \notag \\
	&= \underset{\g,\x}{\arg\min}~ \norm{\y-\P\x}_2^2 + \lambda \sum_i \left[ z_i \gamma_i + \left( x_i^2 + 2 b_i \right) \gamma_i^{-1} \right] \notag
\end{align}

Fixing $\x$ and minimizing with respect to $\g$ by setting the derivative to zero gives
\begin{equation}
	\gamma_i^* = z_i^{-1/2} \sqrt{x_i^2 + 2 b_i} \label{eq:sbldf-rwl1-g}
\end{equation}
where we have taken the positive component of the radical because as variances, each $\gamma_i$ must be positive.

Substituting $\g^{(k+1)}$ from (\ref{eq:sbldf-rwl1-g}) and minimizing with respect to $\x$ gives
\begin{align}
\x^{(k+1)} &= \underset{\left\{x_i \colon \gamma_i \geq \tau \right\}}{\arg\min}~ \norm{\y-\P\x}_2^2 + 2 \lambda \sum_i z_i^{1/2} \sqrt{x_i^2 + 2 b_i} \label{eq:sbldf-rwl1-x}
\end{align}

The complete majorization-minimization procedure alternates between updating the majorizer using (\ref{eq:sbldf-rwl1-z}) and (\ref{eq:sbldf-rwl1-g}), and minimizing it using (\ref{eq:sbldf-rwl1-x}). (Note that $\Sy$ depends on $\g$ and must be updated at each iteration.) For numerical stability, a pruning rule similar to the one used in static SBL is used in the minimization step: when a $\gamma_i$ becomes lower than some threshold $\tau$, the corresponding $x_i$ is fixed to zero and dictionary column $\bm{\phi}_i$ is removed.

\section{Details of Fast Marginal Likelihood}
\label{sec:appendix-fml}

\subsection{Calculation of stationary points of $\ell(\gamma_j)$}

In this section, we give expressions for computing the roots of the cubic expression in (\ref{eq:sbl-fml-lgammaj}) which are used to determine the optimal index and action for the next fast marginal likelihood iteration.

Defining constants $\Delta_0 = c_b^2 - 3c_ac_c$, $\Delta_1 = 2c_b^3 - 9c_ac_bc_c + 27c_a^2c_d$, and determinant $\Delta = 18c_a c_b c_c c_d - 4c_b^3c_d + c_b^2c_c^2 - 4c_a c_c^3 - 27c_a^2 c_d^2$, the roots can be shown to be
\begin{equation*}
\widetilde{\gamma}_j = \begin{cases}
\left\{ -\frac{1}{3c_a}\left( c_b + \zeta^k C + \frac{\Delta_0}{\zeta^k C} \right) \right\}_{k=0,1,2}, & \Delta \neq 0,~ C \neq 0 \\
\left\{ \frac{9 c_a c_d - c_b c_c}{2 \Delta_0}, \frac{4 c_a c_b c_c - 9 c_a^2 c_d - c_b^3}{c_a \Delta_0} \right\}, & \Delta = 0,~ \Delta_0 \neq 0 \\
-\frac{c_b}{3 c_a}, & \Delta = 0,~ \Delta_0 = 0, \\
\end{cases}
\end{equation*}
where $\zeta = -\frac{\sqrt{3}-1}{2}\sqrt{-1}$ and $C = \sqrt[3]{\frac{1}{2} \left( \Delta_1 \pm \sqrt{\Delta_1^2 - 4 \Delta_0^3} \right)}$. These three cases correspond to when there are non-repeated roots, a double root and a single root, and a triple root, respectively.

We can also find the roots as the eigenvalues of the companion matrix
\begin{equation*}
	\bm{C} = \left[ \begin{array}{ccc} 0 & 0 & -\frac{c_0}{c_3} \\ 1 & 0 & -\frac{c_1}{c_3} \\ 0 & 1 & -\frac{c_2}{c_3} \end{array} \right].
\end{equation*}
We have empirically found that this method is more numerically stable than the analytic expressions. See \cite[Sec.~7.4.6]{golub2013matrix} and references therein for more details on the companion matrix method and its stability.

\subsection{Efficient updates for fast marginal likelihood}

The following expressions efficiently implement the \emph{re-estimate}, \emph{add}, and \emph{delete} actions to element $j$ in the fast marginal likelihood algorithm \cite{tipping2003fast}.

\subsubsection{Re-estimating an element}

Defining $\kappa_j = \left[ \left( \widetilde{\gamma_j}^{-1} - \gamma_j^{-1} \right)^{-1} + \left( \Sxy \right)_{jj} \right]^{-1}$ and repeatedly applying the matrix inversion lemma yields
\begin{align*}
	\Sxy^{(\mathrm{new})} &= \Sxy - \kappa_j \Sxy_j \Sxy_j^T, \\
	\muxy^{(\mathrm{new})} &= \muxy - \kappa_j \muxy_i \Sxy_j, \\
	S_i^{(\mathrm{new})} &= S_i + \kappa_j \left( \noisevar \Sxy_j^T \P^T \bm{\phi}_i \right)^2, \label{eq:sbl-fml-efficient-reest-S} \\
	Q_i^{(\mathrm{new})} &= Q_i + \kappa_j \muxy_j \noisevar \Sxy_j^T \P^T \bm{\phi}_i.
\end{align*}

\subsubsection{Adding an element}

Defining $\omega_j = \left(\widetilde{\gamma_j}^{-1}+S_j\right)^{-1}$, using the block matrix inversion identity \cite[Sec.~9.1.3]{petersen2012matrix} and applying the matrix inversion lemma yields
\begin{align*}
	\Sxy^{(\mathrm{new})} &= \left[ \begin{array}{cc} \Sxy + \noisevar^2 \omega_j \Sxy \P^T \bm{\phi}_j \bm{\phi}_j^T \P \Sxy & - \noisevar \omega_j \Sxy \P^T \bm{\phi}_j \\ -\noisevar \omega_j \bm{\phi}_j^T \P \Sxy & \omega_j \end{array} \right], \\
	\muxy^{(\mathrm{new})} &= \left[ \begin{array}{c} \muxy + \noisevar \omega_j Q_j \Sxy \P^T \bm{\phi}_j \\ \omega_j Q_j \end{array} \right], \\
	S_i^{(\mathrm{new})} &= S_i - \omega_j \left[ \noisevar \bm{\phi}_i^T \bm{\phi}_j - \noisevar^2 \bm{\phi}_i^T \P \Sxy \P^T \bm{\phi}_j \right]^2, \\
	Q_i^{(\mathrm{new})} &= Q_i - \omega_j Q_j \left[ \noisevar \bm{\phi}_i^T \bm{\phi}_j - \noisevar^2 \bm{\phi}_i^T \P \Sxy \P^T \bm{\phi}_j \right]. 
\end{align*}

\subsubsection{Deleting an element}

By setting $\widetilde{\gamma_j} = 0$, we have $\kappa_j = \Sxy_{jj}^{-1}$, and the deletion rules follow easily from the re-estimation rules:

\begin{align*}
	\Sxy^{(\mathrm{new})} &= \Sxy - \frac{1}{\Sxy_{jj}} \Sxy_j \left(\Sxy\right)_j^T,\\
	\muxy^{(\mathrm{new})} &= \muxy - \frac{1}{\Sxy_{jj}} \Sxy_j \Sxy_j^T, \\
	S_i^{(\mathrm{new})} &= S_i + \frac{1}{\Sxy_{jj}} \left( \noisevar \Sxy_j^T \P^T \bm{\phi}_i \right)^2, \\
	Q_i^{(\mathrm{new})} &= Q_i + \frac{1}{\Sxy_{jj}} \muxy_j \left(\noisevar \Sxy_j^T \P^T \bm{\phi}_i \right).
\end{align*}

%

\end{document}